\documentclass[journal,10pt,twocolumns,twoside]{IEEEtran}
\IEEEoverridecommandlockouts
\usepackage{algorithm}
\usepackage{algpseudocode}
\usepackage{multirow}
\usepackage{amsthm}

\newcommand\hmmax{0}
\newcommand\bmmax{0}

\usepackage{dsfont}
\usepackage{lipsum}
\usepackage{graphicx}
\usepackage{tikz}
\usetikzlibrary{arrows.meta, positioning, shapes.geometric}
\usetikzlibrary{calc}
\usepackage{pgfplots}
\usepackage{xcolor}
\usepackage{color, colortbl}
\usepackage{amsmath}
\usepackage{bbm}
\usepackage{amsfonts, amssymb}
\usepackage{mathtools}
\usepackage{cite}
\usepackage[nolist]{acronym}
\usepackage{balance}

\usepackage{booktabs} 		% cute tables
\usepackage{diagbox}		% diagonal cell division in table
\usepackage{upgreek}
\usepackage{bbm}
\usepackage{Files/bm}
\usepackage{Files/winsnotation}
\usepackage{setspace}	 	% stretch on algorithm2e
\usepackage{comment}
\usepackage{array}
\newcolumntype{C}[1]{>{\centering\arraybackslash}p{#1}}

\makeatletter
\def\endthebibliography{%
  \def\@noitemerr{\@latex@warning{Empty `thebibliography' environment}}%
  \endlist
}
\makeatother
\IEEEoverridecommandlockouts
\usepackage{amsthm}
\theoremstyle{plain}
\newtheorem{theorem}{Theorem}
\theoremstyle{definition}

\newtheorem{definition}{Definition}
\newtheorem{remark}{Remark}
\newtheorem{lemma}{Lemma}

\pgfplotsset{compat=1.17}
\begin{document}

\title{Design and Performance Analysis of High-Rate GLDPC Codes for Optical Communications}

\author{%
    \IEEEauthorblockN{Alessandro Mirri,~\IEEEmembership{Graduate~Student~Member,~IEEE,} 
    Lorenzo Valentini,~\IEEEmembership{Member,~IEEE,}\\
    Enrico Paolini,~\IEEEmembership{Senior~Member,~IEEE}}
    \thanks{Alessandro Mirri, Lorenzo Valentini, and Enrico Paolini are with CNIT/WiLab, DEI, University of Bologna, Italy. Email: \{alessandro.mirri7, lorenzo.valentini13, e.paolini\}@unibo.it.}
}

% make the title area
% Don't write page number 0 to the cover page.
\maketitle 

\begin{acronym}
% usage: \ac{SW}, \acp{SW} for plurals \acf{SW} Use the full name of the acronym.
%\acs{SW}Use the acronym, even before the first corresponding \ac command
%\acl{acronym}Expand the acronym without using the acronym itself.
\acro{AWGN}{additive white Gaussian noise}
\acro{APP}{a posteriori probability}
\acro{BDD}{bounded distance decoding}
\acro{bi-AWGN}{binary-input additive white Gaussian noise}
\acro{BCH}{Bose-Chaudhuri-Hocquenghem}
\acro{BER}{bit error rate}
\acro{BPSK}{binary phase shift keying}
\acro{BSC}{binary symmetric channel}
\acro{CER}{codeword error rate}
\acro{CN}{check node}
\acro{DE}{density evolution}
\acro{eBCH}{extended BCH}
\acro{eH}{extended Hamming}
\acro{eRBF}{enhanced reliability bit flipping}
\acro{FEC}{forward error correction}
\acro{GLDPC}{generalized LDPC}
\acro{HD}{hard-decision}
\acro{HDD}{hard-decision decoding}
%\acro{HOSC}{high-order spatially-coupled}
\acro{HOSC}{higher-order staircase code}
\acro{iBDD}{iterative BDD}
\acro{iBDD-RF}{iBDD with random flipping}
\acro{i.i.d.}{independent and identically distributed}
\acro{OH}{overhead}
\acro{TB-iBDD}{threshold-based iBDD-SR}
\acro{TB-iBDD-SR}{threshold-based IBDD-SR}
\acro{iBDD-SR}{iBDD with scaled reliability}
\acro{iBDD-TBF}{iterative BDD with threshold-based flipping}
\acro{LDPC}{low-density parity-check}
\acro{LLR}{log likelihood ratio}
\acro{PEG}{progressive edge growth}
\acro{QC}{quasi-cyclic}
\acro{RBF}{reliability bit flipping}
\acro{RS}{Reed-Solomon}
\acro{NCG}{net coding gain}
\acro{SC}{staircase code}
\acro{SD}{soft-decision}
\acro{SDD}{soft-decision decoding}
\acro{SNR}{signal-to-noise ratio}
\acro{SPC}{single parity-check}
\acro{SRAA}{shift-register-adder-accumulator}
\acro{SVN}{super variable node}
\acro{TB-BMP}{threshold-based binary message passing}
\acro{VN}{variable node}
\end{acronym}
\setcounter{page}{1}

\begin{abstract}
Forward error correction is a key enabler of modern optical communication systems, where extremely low bit error rate (BER), high throughput, and stringent latency constraints must be simultaneously satisfied.
In this work, we investigate the design of high-rate generalized low-density parity-check (GLDPC) codes for next-generation optical links, with particular focus on moderate block lengths and practical hard decision decoding.
High-rate GLDPC codes typically rely on short component codes and low variable node degrees, which may significantly degrade the error floor.
To address this issue, we develop a design methodology that balances code rate and low BER performance through the joint selection of structural code parameters.
An analytical framework is presented to predict the waterfall behavior under an explicit independence assumption and to characterize the finite-length error floor region. 
%characterize both the asymptotic waterfall behavior and the finite-length error floor region.
Numerical results show that properly designed high-rate GLDPC codes achieve favorable performance-complexity tradeoffs, making them attractive candidates for emerging optical systems requiring low BER and low latency transmission at very high data rates.
\end{abstract}

%%%%%%%%%%%%%%%%%%%%%%%%%%%%%%%%%%%%%%%%%%%%%%%%%%%%%%
%%%%%%%%%%%%%%%%%%%%%%%%%%%%%%%%%%%%%%%%%%%%%%%%%%%%%%
\section{Introduction}
\label{sec:Introduction}

The ever growing optical communication traffic is driving the need for increasingly efficient \ac{FEC} schemes capable of operating at extremely high data rates while guaranteeing very low output \ac{BER}.
Emerging standards such as $400$~GbE, $800$~GbE and future $1.6$~TbE impose stringent constraints not only on reliability, but also on latency, power consumption, hardware complexity, and coding overhead~\cite{Justensen2010:FEC_OTN}.
Therefore, as throughput scales upward, the design of practical \ac{FEC} solutions becomes increasingly challenging.

A first class of solutions is based on end-to-end \ac{FEC}, where a single code protects the entire communication link.
These schemes are particularly attractive for short-reach and high-throughput interconnects because they preserve a simple transceiver architecture, leading to very low latency.
Representative examples are the standardized high-rate shortened \ac{RS} block codes KR4 (\ac{RS}$(528,514)$) and KP4 (\ac{RS}$(544,514)$), widely adopted in Ethernet systems~\cite{IEEE8023bj}.
Their relatively short block lengths enable compact implementations and very fast \ac{HD} decoding, although the achievable \ac{NCG} remains limited.
To improve performance while retaining \ac{HD} decoding, more advanced schemes relying on product codes~\cite{Elias1954:ProductCodes} have been proposed~\cite{Blaum1988:ArrayCodes,fan2001array}.
Prominent examples are \acp{SC}~\cite{Smith2012:StaircaseCodes}, where algebraic component codes are iteratively decoded over a sliding window of coupled blocks, and \acp{HOSC}~\cite{Shehadeh2025:HOSC}, which extend the staircase principle through stronger or higher-dimensional component code coupling.
Another example of end-to-end code is OFEC~\cite{Wang2023:CFEC_OFEC,OpenROADM2023}, where a limited number of \ac{SD} iterations is first performed, followed by subsequent \ac{HD} stages in order to reduce complexity.
% while preserving high \ac{NCG}.
These schemes achieve significantly larger \ac{NCG} than conventional block \ac{FEC} and enable efficient encoder implementations based on component codes processing~\cite{Li2020:EncoderSC}.
However, such gains are obtained at the expense of increased decoding latency due to the window-based architecture.

A second class of approaches relies on segmented, hybrid, or concatenated architectures, where different codes are employed in different portions of the link or across multiple serial decoding stages~\cite{Cai2019:Concatenated_Staircase_Hamming_FEC}.
This paradigm is common in long-haul coherent optical systems, where a powerful inner code is combined with a lower-complexity outer code to eliminate residual errors.
Representative examples include CFEC schemes~\cite{Wang2023:CFEC_OFEC}, where an \ac{SD} inner \ac{eH} code is concatenated with an outer \ac{SC} with algebraic \ac{BCH} component codes, thereby combining strong waterfall performance with error floor reduction.
Further examples are concatenated extensions of KP4, such as KP4 combined with \ac{SPC} outer code~\cite{Lentner2023:ConcatenatedKp4}, or KP4 combined with \ac{eH} outer code~\cite{Nedelcu2022:KP4_eH}.
Although these hybrid and concatenated solutions provide excellent \ac{NCG}, they come at the expense of increased implementation complexity, buffering requirements, and decoding latency due to the presence of mixed \ac{HD}/\ac{SD} decoding stages.
This motivates the search for alternative coding schemes capable of approaching state-of-the-art performance while preserving the low latency and high throughput of purely \ac{HD} architectures.

While \ac{HD} decoding significantly reduces internal data flow and implementation complexity compared with \ac{SD} architectures, it typically suffers a performance loss.
At the same time, high-speed optical systems operate at large pre-\ac{FEC} error rates, thereby requiring strong \ac{NCG} to meet stringent output targets such as $\text{BER}^{\star}=10^{-14}$ or $10^{-15}$.
These opposing trends expose a central challenge for next-generation optical links: achieving higher coding performance while simultaneously reducing decoding complexity and latency.
In this context, high-rate \ac{GLDPC} codes represent a compelling alternative.
By replacing simple parity constraints with stronger component codes, \ac{GLDPC} constructions can provide powerful \ac{HD} decoding gains while retaining sparse graph structures and efficient iterative decoders~\cite{Tanner1981:GLDPC,sipser2002expander}.
Their flexibility makes them attractive for optical systems requiring very small overhead and very high throughput~\cite{Djordjevic2005:GLDPC_Optical,Djordjevic2008:GLDPC_RM_Optical,Liu2018:GLDPC_URLLC}.
However, the operating regime considered in this work remains particularly challenging, as it combines stringent \ac{NCG} and overhead requirements, very-low \ac{BER} targets, and latency constraints that have not been jointly addressed in previous \ac{GLDPC}-based optical communication studies.

Stringent overhead constraints naturally push the design toward very high-rate codes, which are often associated with short component codes and low \ac{VN} degrees.
Concurrently, practical implementation and latency considerations discourage extremely long code constructions typically employed in product-like or staircase-based schemes.
In this work, we therefore focus on a shorter-length high-rate regime that remains attractive from both implementation and latency perspectives, while still achieving low \ac{BER} performance.
Despite its practical relevance, this design space remains comparatively underexplored.
The considered regime also introduces important challenges from the decoding perspective.
The proposed codes are generally more sensitive to harmful local graph configurations, often referred to as trapping sets or stalling sets, which cannot be resolved by conventional \ac{BDD}~\cite{Miladinovic2008:GLDPC,Chilappagari2010:TrappingGLDPC}.
These events typically dominate the error floor and may prevent compliance with the very low BER targets required in optical standards. 
Consequently, the design of high-rate \ac{GLDPC} codes cannot rely solely on waterfall-oriented analytical predictions, but must jointly account for finite-length error floor behavior.
Motivated by these considerations, we investigate the design and performance analysis of high-rate \ac{GLDPC} codes that simultaneously offer good waterfall performance, low error floors, and reduced decoding complexity.

The main contributions of this paper are summarized as follows:
\begin{itemize}
    \item We investigate a practically relevant and underexplored operating regime for \ac{GLDPC} codes in optical \ac{FEC}, namely high-rate configurations where \ac{NCG}, very low \ac{BER}, and latency constraints must be jointly satisfied.

    \item We propose a \ac{GLDPC} code design methodology that balances key structural parameters, including the component codes choice and the \acp{VN} degree, in order to preserve high-rate operation while mitigating harmful graph substructures that stall the decoder.

    \item We derive a finite-length error floor analysis that predicts the dominant rare-event error patterns under miscorrection-free decoding and enables performance estimation in the low \ac{BER} regime.
    
    \item We develop an approximate analytical framework, based on an explicit independence assumption, to predict the waterfall performance under \ac{HD} decoding and provide design insights on threshold behavior. 

    \item We validate the proposed methodology through extensive Monte Carlo simulations and comparisons with reference \ac{FEC} schemes, highlighting favorable performance-complexity tradeoffs.
\end{itemize}

The remainder of the paper is organized as follows.
Section~\ref{sec:Preliminaries} provides preliminaries on \ac{GLDPC} codes and \ac{iBDD}, together with the adopted channel model.
Section~\ref{sec:Design} introduces the proposed high-rate \ac{GLDPC} codes design.
Section~\ref{sec:ErrorFloorAnalysis} develops the finite-length error floor estimation under miscorrection-free decoding.
Section~\ref{sec:AsymptoticWaterfallAnalysis} presents the approximate waterfall analysis.
Section~\ref{sec:NumericalResults} reports numerical results, and Section~\ref{sec:Conclusions} concludes the paper.

\emph{Notation:} Bold uppercase and lowercase letters denote matrices and vectors, respectively.
The $i$-th entry of a vector $\V{v}$ is denoted by ${v}_i$.
For a set $\mathcal{S}$, $|\mathcal{S}|$ denotes its cardinality.
The floor and ceiling operators are denoted by $\lfloor\cdot\rfloor$ and $\lceil\cdot\rceil$, respectively, and function composition by $\circ$.
The probability of an event $\mathcal{E}$ is written as $\mathbb{P}(\mathcal{E})$.
The indicator function is denoted by $\mathbbm{1}\{\cdot\}$.

%%%%%%%%%%%%%%%%%%%%%%%%%%%%%%%%%%%%%%%%%%%%%%%%%%%%%%
%%%%%%%%%%%%%%%%%%%%%%%%%%%%%%%%%%%%%%%%%%%%%%%%%%%%%%

%%%%%%%%%%%%%%%%%%%%%%%%%%%%%%%%%%%%%%%%%%%%%%%%%%%%%%
%%%%%%%%%%%%%%%%%%%%%%%%%%%%%%%%%%%%%%%%%%%%%%%%%%%%%%
\section{Preliminaries and Channel Model}
\label{sec:Preliminaries}

\begin{figure}[t]
    \centering
    \includegraphics[width=0.85\columnwidth]{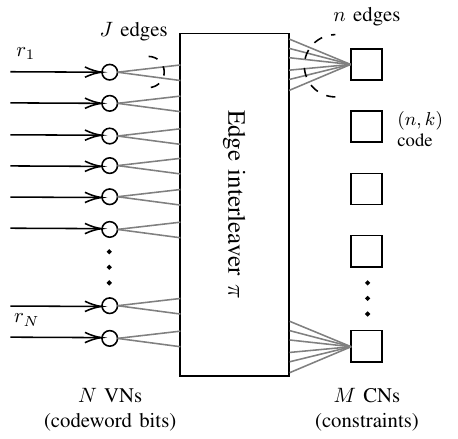}
    \caption{Graphical representation of a regular GLDPC code: $N$ VNs connected through an edge interleaver $\pi$ to $M$ CNs implementing $(n,k)$ component codes.}
    \label{fig:gldpc_structure}
\end{figure}
\Ac{GLDPC} codes are a class of \ac{FEC} codes that admit a natural graphical representation through their Tanner graph \cite{Tanner1981:GLDPC,Lentmaier1999:GLDPC_Hamming,Simegn2025:LowComplexGLDPC}.
A \ac{GLDPC} code is defined by a bipartite graph composed of $N$ \acp{VN} and $M$ \acp{CN}.
A GLDPC codeword is denoted by $\V{c} \in \mathbb{F}_2^N$, where $c_1,\dots,c_N$ are the encoded bits.
The \acp{CN} enforce local constraints on subsets of $\V{c}$.
Each \ac{CN} is associated with a component code $\mathcal{C}$, typically a linear block code, with parameters $(n,k,d_{\min})$ or equivalently $(n,k,t)$, where $t = \left\lfloor \frac{d_{\min}-1}{2} \right\rfloor$ denotes its error-correction capability.
Thus, each \ac{CN} imposes $n-k$ parity constraints on its neighboring \acp{VN}. 
This construction generalizes standard \ac{LDPC} codes~\cite{Gallager1962:LDPC}, recovered when each \ac{CN} enforces an \ac{SPC} constraint.

The connectivity between \acp{VN} and \acp{CN} is defined by an \emph{edge interleaver}, which specifies the permutation pattern of the graph edges and ultimately defines the code structure and its cycle properties.
A \ac{GLDPC} code is called \emph{regular} if all \acp{VN} have degree $J$ and all \acp{CN} have degree $n$ (see Fig.~\ref{fig:gldpc_structure}). 
It is \emph{irregular} if \ac{VN} degrees vary and/or different \acp{CN} employ different component codes (also referred to as \emph{hybrid} \ac{GLDPC} constructions).
Efficient encoding of GLDPC codes is often based on protograph-based constructions~\cite{Liva2008:QuasiCyclicLDPC,smarandache2025generalizedquasicyclicldpccodes,smarandache2025generalizedquasicyclicldpccodes_ISIT}, where a small base graph is lifted to generate a large structured Tanner graph. The resulting \ac{QC} structure enables efficient encoder implementations~\cite{Li2006:EncodingQCLDPC}.

\subsection{Channel Model}
\label{subsec:ChannelModel}
We assume transmission over a \ac{bi-AWGN} channel with \ac{BPSK} modulation, as commonly adopted in optical communication systems.
The received vector is denoted by $\V{r} \in \mathbb{R}^N$, and its corresponding \ac{HD} version by $\V{y} \in \mathbb{F}_2^N$.
The channel output for the $i$-th coded bit $c_i$ is given by
\begin{align}
    r_i = x_i + z_i,
\end{align}
where $x_i = 1 - 2\,c_i$ maps the binary bit $c_i \in \{0,1\}$ to $\{+1,-1\}$, and $z_i \sim \mathcal{N}(0,\sigma_\mathrm{n}^2)$ is a sample of Gaussian noise.
The noise variance is
\begin{align}
\sigma_\mathrm{n}^2 = \frac{1}{2\, R \, E_\mathrm{b}/N_0}\,,
\end{align}
where $R$ is the \ac{GLDPC} code rate and $E_\mathrm{b}/N_0$ is the \ac{SNR}.
The \ac{HD} vector is defined as $y_i = \mathbbm{1}\{r_i < 0\}$, for all $i=1,\dots,N$, and $\V{y} = (y_1,\dots,y_N)$.
A vector $\V{y}$ corresponds to a valid \ac{GLDPC} codeword if and only if all \acp{CN} recognize the corresponding local patterns as valid codewords of their associated $(n,k,t)$ component codes.

\subsection{Decoding of GLDPC Codes}
\label{subsec:GLDPC_Decoding}
Iterative decoding of \ac{GLDPC} codes can be based on either message-passing or bit-flipping strategies.
Message-passing algorithms rely on the exchange of extrinsic information between \acp{VN} and \acp{CN}, where messages may be hard values~\cite{Chase1972:DecBlockCodes, Jian2017:IterativeHDD} or soft values~\cite{Pyndiah1998:DecProd}.
Bit-flipping decoding instead updates the values of selected \acp{VN} according to local decisions made at the \acp{CN}~\cite{sipser2002expander,Chilappagari2010:TrappingGLDPC}.
We focus on \ac{HD} bit-flipping decoding, referred to as \ac{iBDD}.
At each iteration, \ac{BDD} is sequentially and independently applied at every \ac{CN} to the current estimates of the neighboring \acp{VN}. Whenever decoding is successful, the corrected component code bits are written back to the associated \acp{VN}, thereby updating the global \ac{GLDPC} codeword estimate for subsequent decoding steps.
It can be interpreted as a generalized bit-flipping procedure where each \ac{CN} enforces its associated component code constraint via \ac{BDD}~\cite{Burshtein2025:GLDPC_BDD}.

Let $\mathcal{N}_{\mathrm{VN}(i)}$ and $\mathcal{N}_{\mathrm{CN}(j)}$ denote the neighborhoods of VN $i$ and CN $j$, respectively. Let $\ell_{\max}$ be the maximum number of iterations.
The decoder operates on \acp{HD} only.
Initialization is given by $c_i^{0} = y_i$ for all $i=1,\dots,N$.
At iteration $\ell=1,\dots,\ell_{\max}$, the \acp{CN} are processed sequentially.
Whenever \ac{BDD} at a given \ac{CN} succeeds, the corresponding \acp{VN} are immediately updated, such that subsequent \acp{CN} may exploit the newly corrected values.
For each \ac{CN} $j$, the decoder forms the received vector $\V{c}_{\mathcal{N}_{\mathrm{CN}(j)}}^{\ell-1} = \big( c_i^{\ell-1} : i \in \mathcal{N}_{\mathrm{CN}(j)} \big)$ and applies \ac{BDD}.
The decoder outputs the component code codeword $\hat{\V{c}}^{\ell}(j) = \mathrm{BDD}(\V{c}_{\mathcal{N}_{\mathrm{CN}(j)}}^{\ell-1})$.
The bounded-distance decoder corrects all error patterns of Hamming weight up to $t=\left\lfloor \frac{ d_\mathrm{min}-1}{2}\right\rfloor$, where $d_\mathrm{min}$ is the minimum distance of the component code.
If the vector $\V{c}_{\mathcal{N}_{\mathrm{CN}(j)}}^{\ell-1}$ lies within the radius $t$ of the transmitted component code codeword, correct decoding is guaranteed.
If another codeword lies within the decoding radius, a miscorrection may occur; otherwise, decoding fails and no update is performed.

After all \ac{CN} updates, the new global estimate of the \ac{GLDPC} codeword $\V{c}^{\ell}$ is obtained.
At termination, the decoder checks whether all \ac{CN} constraints are satisfied, i.e., whether each \ac{CN} input corresponds to a valid codeword of its associated component code.
If so, $\V{c}^{\ell} = \hat{\V{c}}$ is declared a valid \ac{GLDPC} codeword; otherwise, a decoding failure is declared.
The procedure is summarized in Algorithm~\ref{alg:iBDD}.
\begin{algorithm}[t]
\caption{iBDD for GLDPC Codes}
\label{alg:iBDD}
\begin{algorithmic}[1]
\State \textbf{Input:} Hard decisions $\V{y} \in \mathbb{F}_2^N$
\State \textbf{Init:} $c_i^{0} \leftarrow y_i$, \; $i=1,\dots,N$
\For{$\ell = 1$ to $\ell_{\max}$}
    \For{$j = 1$ to $M$}
        \State $\V{c}_{\mathcal{N}_{\mathrm{CN}(j)}}^{(\ell-1)} = \big( c_i^{\ell-1} : i \in \mathcal{N}_{\mathrm{CN}(j)} \big)$
        \State $\hat{\V{c}}^{\ell}(j) = \mathrm{BDD}(\V{c}_{\mathcal{N}_{\mathrm{CN}(j)}}^{(\ell-1)})$
        \If{BDD succeeds}
            % \State Replace the entries of $\V{c}^{(\ell)}$
            % corresponding to $\mathcal{N}_{\mathrm{CN}(j)}$
            % with $\hat{\V{c}}^{\ell}(j)$
            \State \parbox[t]{\dimexpr\linewidth-5em}{Replace the entries of $\V{c}^{(\ell)}$
            corresponding to $\mathcal{N}_{\mathrm{CN}(j)}$
            with $\hat{\V{c}}^{\ell}(j)$}
        \EndIf
    \EndFor
\EndFor
\If{all CN constraints are satisfied} 
    \State \textbf{Output:} GLDPC codeword $\hat{\V{c}} \in \mathbb{F}_2^N$
\Else
    \State \textbf{Output:} decoding failure
\EndIf
\end{algorithmic}
\end{algorithm}

The performance of \ac{iBDD} can be further improved by exploiting limited soft information from the channel, such as the \acp{LLR}.
A notable example is \ac{iBDD-SR}~\cite{Liva2019:BinaryMessagePassing}, originally proposed for product codes, and readily extendable to \ac{GLDPC} codes.
\ac{iBDD-SR} enhances conventional \ac{iBDD} by incorporating channel reliability information into the decoding process.
Specifically, each component decoder combines the binary output of the \ac{BDD} with an iteration-dependent reliability weight and the channel \acp{LLR}.
Although the \acp{VN} updates remain binary, the internal use of these soft metrics allows the decoder to refine hard decisions more effectively than standard \ac{iBDD}.
This soft-assisted update yields improved performance while maintaining a complexity close to that of conventional \ac{iBDD}.

A related decoder, the \ac{iBDD-RF} algorithm proposed in~\cite{Li2025:IterativeRandomBDD}, builds upon the same fundamental principle as \ac{iBDD-SR}, namely the formation of a soft reliability metric obtained from a weighted combination of the \ac{BDD} decision and the channel \ac{LLR}. 
However, \ac{iBDD-RF} introduces a stochastic mechanism to help the decoder escape from unfavorable decoding states. 
Specifically, when the magnitude of the reliability value falls below a predefined threshold, the corresponding bit is randomly flipped with a certain probability. 
This random perturbation can help break harmful trapping or stalling configurations that would otherwise prevent further decoding progress.

Another enhancement is proposed in~\cite{zhao2024:ThresholdBasedMessagePassing} through the \ac{TB-BMP} algorithm, which replaces random flipping with a deterministic threshold-based mechanism augmented with memory. 
In \ac{TB-BMP}, the reliability associated with each \ac{VN} depends not only on the outcome of the current \ac{BDD} operation, but also on a stored memory value and a predefined reliability threshold. 
This leads to a history-aware flipping strategy that improves the decoder's ability to escape persistent error patterns.

In the remainder of this paper, we adopt \ac{iBDD} as the reference decoding algorithm. 
This choice is motivated by its low-complexity \ac{HD} nature, analytical tractability, and practical relevance for high-throughput optical implementations, while fully SD decoding based on component code a posteriori probability decoding and graph-level belief propagation entails additional processing, memory, and interconnect requirements.
Since the main focus of this work is the design of high-rate \ac{GLDPC} codes and the characterization of their performance under low-complexity decoding, the use of \ac{iBDD} provides a meaningful and consistent benchmark. 
We emphasize that the use of \ac{iBDD} is not intended as a novel decoding contribution, but rather as a deliberate choice that defines the decoding regime considered in this work.
The proposed code constructions can also be considered with more advanced decoding schemes, such as \ac{iBDD-SR}, \ac{iBDD-RF}, \ac{TB-BMP}, or other reliability-aided extensions. 
Their analysis and performance evaluation under these alternative decoders are beyond the scope of this work.

%%%%%%%%%%%%%%%%%%%%%%%%%%%%%%%%%%%%%%%%%%%%%%%%%%%%%%
%%%%%%%%%%%%%%%%%%%%%%%%%%%%%%%%%%%%%%%%%%%%%%%%%%%%%%

%%%%%%%%%%%%%%%%%%%%%%%%%%%%%%%%%%%%%%%%%%%%%%%%%%%%%%
%%%%%%%%%%%%%%%%%%%%%%%%%%%%%%%%%%%%%%%%%%%%%%%%%%%%%%
\section{Design of Fair-Regular GLDPC Codes}
\label{sec:Design}

We consider a regular \ac{GLDPC} code with $N$ \acp{VN} and $M$ \acp{CN}. 
Each \ac{CN} is connected to $n$ \acp{VN}, and each \ac{VN} is connected to $J$ \acp{CN}. 
Consequently, the total number of edges in the Tanner graph satisfies
\begin{align}
M n = J N\,.
\end{align}
The nominal code rate $R$ is determined by the rates of the component codes associated with the \acp{CN}.
For a regular \ac{GLDPC} code employing a single component code class of dimension $k$, the nominal rate is
\begin{align}
R = \frac{K}{N} = 1 - \frac{M(n-k)}{N} = 1 - J(1-R_c),
\end{align}
where $K$ is the number of information bits of the \ac{GLDPC} code and $R_c = k/n$ is the component code rate.
For component codes such as binary \ac{BCH} codes with blocklength $n$ and error-correction capability $t$, the code rate satisfies
\begin{align}
R_c = 1 - \frac{p}{n} \ge 1 - \frac{mt}{n} \,, 
\end{align}
where $p = n-k \le mt$ denotes the number of parity-check equations of the component code and $m$ is the smallest positive integer such that $n\le 2^m-1$.
For the primitive \ac{BCH} codes considered in this work, $n=2^m -1$ and the redundancy is $p=mt$, yielding $R_c = 1-(mt/n)$.\footnote{For the corresponding \ac{eBCH} codes, $p = mt + 1$.}
For a regular \ac{GLDPC} code with different component codes classes\footnote{The component codes associated with the \acp{CN} may be heterogeneous (e.g., with different dimensions $k$ and error-correction capabilities $t$), provided that they share the same blocklength $n$.}, the nominal rate becomes
\begin{align}
R
=1-\frac{\sum_{g=1}^{G} M_g(n-k_g)}{N}
=1-\frac{1}{N}\sum_{g=1}^{G} M_g p_g,
\end{align}
where $M_g$ denotes the number of \acp{CN} associated with the class-$g$ component code ($\sum_{g=1}^{G} M_g = M$, with $G$ distinct component code classes), and $p_g=n-k_g$ is the number of parity constraints contributed by a class-$g$ component code.
The code overhead is defined as
\begin{align}
\text{OH} = \frac{N-K}{K} = \frac{1-R}{R}\,.
\end{align}
\begin{remark}
    The above expressions provide the \emph{nominal} code rate, which assumes that all component code constraints are linearly independent.
    The actual code rate is computed after constructing the full \ac{GLDPC} parity-check matrix, obtained by combining the Tanner-graph adjacency structure with the parity-check matrices of the component codes~\cite{Tanner1981:GLDPC}.
    In practice, some constraints may be redundant, so the effective rate can be slightly higher than the nominal one, resulting in a marginally lower overhead. 
    Table~\ref{tab:Design_Parameters} reports both nominal and actual rates for the code designs considered in this work.
\end{remark}    

\subsection{Fair-Regular GLDPC Codes Construction}
\label{subsec:FairRegular}
\begin{figure}[t]
    \centering
    % \resizebox{0.99\columnwidth}{!}{        \input{Figures/AdjMatrixPlusBaseCirculantFairRegular_Q4}
    % }
    \includegraphics[width=1.0\columnwidth]{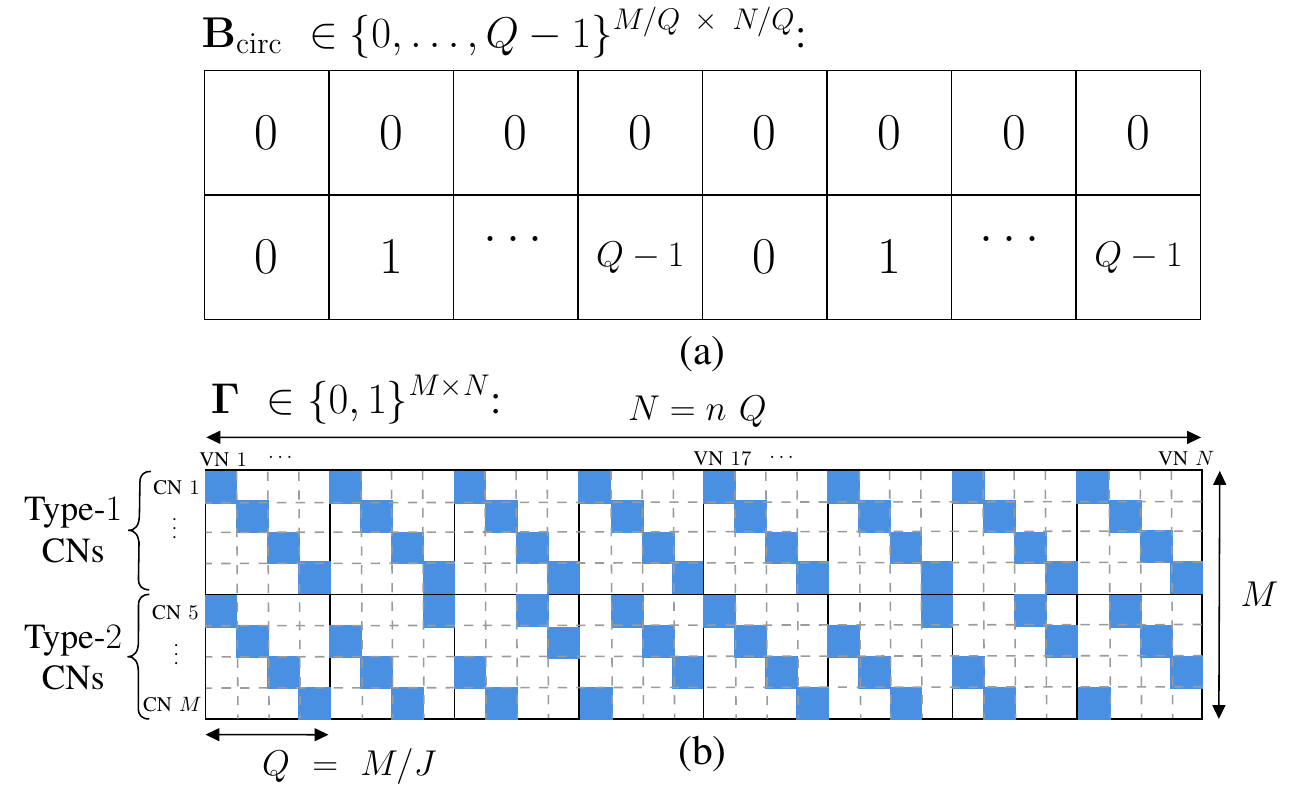}
    \caption{Base matrix and adjacency matrix for the proposed fair-regular \ac{GLDPC} codes construction. Example with $M=8$, $N=32$, $Q=4$ and $J=2$; component code: \ac{eH} ($n=8$). (a) Base circulant matrix $\M{B}_\mathrm{circ}$, whose entries define the circulant shifts. (b) Corresponding lifted adjacency matrix $\M{\Gamma}$ obtained via circulant expansion. Blue squares denote entries equal to $1$, i.e., edges between \acp{CN} and \acp{VN}.}
    \label{fig:AdjMatrixFairRegular}
\end{figure}

We focus on a subclass of regular \ac{GLDPC} codes characterized by a highly structured and uniform connectivity pattern.
In particular, all \acp{VN} have the same degree $J$, all \acp{CN} are associated with component codes of identical length $n$, and the adjacency matrix is designed to enforce a uniform distribution of edges across different \acp{CN}.
Specifically, the $M$ \acp{CN} are partitioned into $J$ disjoint types, each containing $Q = M/J$ nodes.
Each \ac{VN} is connected to exactly one \ac{CN} from each type, ensuring a uniform spreading of connections.
A type-$x$ \ac{CN} is indexed as
\begin{align}
(x-1)\frac{M}{J} + 1 \le j \le x\frac{M}{J} , \quad 1 \le x \le J \,.
\end{align}
Depending on the considered design, all \acp{CN} may employ the same component code or may use different ones, each selected from a set of $G \le J$ classes.
\begin{definition}[Fair-regular GLDPC]
    A \ac{GLDPC} code is said to be \emph{fair-regular} if its adjacency matrix can be partitioned into $J$ groups of $Q \times Q$ circulant permutation submatrices, and if each \ac{VN} is connected to exactly one \ac{CN} from each type.
    Moreover, for any selection of $J$ \acp{CN} containing one \ac{CN} per type, the number of \acp{VN} connected simultaneously to all of them is equal either to zero or to a constant positive integer $v$, where $v$ is independent of the particular selection of \acp{CN}. 
\end{definition}
The proposed fair-regular structure is specifically designed for the highly constrained regime considered in next-generation optical communications, where stringent implementation requirements substantially restrict the admissible \ac{GLDPC} code construction degrees of freedom.
Random and PEG-optimized \ac{GLDPC} constructions are therefore outside the scope of this work: their connectivity is typically optimized to improve waterfall performance, and their inherently unstructured, non-deterministic design does not naturally accommodate the required structural constraints.

The adjacency matrix is constructed deterministically through a protograph-based circulant lifting.
Let $Q$ denote the lifting factor.
The \acp{VN} are partitioned into $N/Q=n$ groups of $Q$ \acp{VN} each, while the \acp{CN} are partitioned into $M/Q=J$ types of $Q$ \acp{CN} each.
The base matrix
\begin{align}
\M{B}_\mathrm{circ} \in \{0,1,\dots,Q-1\}^{M/Q \times N/Q}\,,
\end{align}
specifies the circulant shifts associated with each connection between a \ac{CN} type and a \ac{VN} group.
Each entry of $\M{B}_\mathrm{circ}$ defines a cyclic permutation connecting a group of $Q$ \acp{VN} to a type of $Q$ \acp{CN}.
The first row (row $0$) of $\M{B}_\mathrm{circ}$ contains only zero elements.
For each subsequent row $i \ge 1$, the sequence $0,1,\ldots,Q-1$ is repeated periodically with period $Q^{\,i-1}$, so that each shift value is held constant over $Q^{\,i-1}$ consecutive entries before incrementing.
When the maximum shift value $Q-1$ is reached, the pattern wraps around cyclically and restarts from zero.
The full adjacency matrix $\M{\Gamma} \in \{0,1\}^{M \times N}$ is obtained by expanding each entry of $\M{B}_\mathrm{circ}$ into a $Q \times Q$ circulant permutation block. 
An example of the base matrix and the resulting adjacency matrix for the proposed fair-regular \ac{GLDPC} codes is shown in Fig.~\ref{fig:AdjMatrixFairRegular}.

The fair-regular adjacency matrix structure enables the
construction of a QC block-circulant representation of the global GLDPC parity-check matrix.
By exploiting this structure, encoding can be performed efficiently using the global GLDPC parity-check matrix, even though the component codes share \acp{VN} and therefore cannot be encoded independently.
More specifically, since the adjacency matrix is obtained by circulant lifting of the base matrix $\M{B}_{\mathrm{circ}}$, each entry of $\M{B}_{\mathrm{circ}}$ corresponds to a circulant permutation matrix in the adjacency matrix.
Each row of the adjacency matrix is expanded into the parity-check equations of its component code, using the same component code coordinate assignment for all \acp{CN} of a given type. 
Reordering the resulting rows by \ac{CN} type and parity-check equations yields an array of $Q \times Q$ blocks, each being either a circulant permutation matrix or the null matrix \cite{Liva2008:QuasiCyclicLDPC}.
Related works have also established algebraic approaches for constructing structured generator matrices for QC-\ac{GLDPC} codes, including rank-deficient and mixed-constraint cases~\cite{smarandache2025generalizedquasicyclicldpccodes_ISIT,smarandache2025generalizedquasicyclicldpccodes}.

The array-of-circulant representation of the global parity-check matrix also satisfies the structural conditions required by the hardware-efficient shift-register-based encoding framework presented for \ac{QC}-\ac{LDPC} codes in~\cite{Li2006:EncodingQCLDPC}, which can therefore be adapted to fair-regular \ac{GLDPC} codes.
This framework also covers rank-deficient parity-check matrices \cite[Section~V]{Li2006:EncodingQCLDPC}, as required for the proposed designs.
It provides serial and parallel \ac{SRAA} implementations, including a two-subencoder construction for the rank-deficient case, while its two-stage architecture is developed under the assumptions specified in Section~IV
of~\cite{Li2006:EncodingQCLDPC}.
These architectures exploit the circulant structure and offer tradeoffs between hardware resources and encoding speed.
The lifting factor $Q=32$ used in the considered designs (see Section~\ref{sec:NumericalResults}) gives short shift registers in these SRAA implementations, while the achievable throughput depends on the architecture and clock frequency.
Related hardware encoder implementations are discussed in~\cite{Ren2018:HS_Codec_FPGA,Kang2020:Efficient_FEC_Encoder,Liu2021:LDPCEncoder}.
\begin{remark}
    An important consequence of the proposed fair-regular construction is that \acp{CN} belonging to the same type do not share any \acp{VN}. Therefore, under \ac{iBDD}, all \acp{CN} of a given type can be processed fully in parallel within the same decoding iteration.
\end{remark}
\begin{definition}[Super variable node]
    A \emph{\ac{SVN}} is a non-empty subset of \acp{VN} that share the same set of neighboring \acp{CN}.
    More specifically, a \emph{type-$x$ SVN} is a subset of \acp{VN} such that all \acp{VN} in the subset are connected to the same set of $x$ \acp{CN}, with exactly one \ac{CN} from each of the first $x$ \ac{CN} types, irrespective of their connections to the remaining types.
    Thus, type-$1$ \acp{SVN} partition the \acp{VN} according to their connection to the first \ac{CN} type.
    Each subsequent \ac{CN} type further refines these groups according to the corresponding \ac{CN} connections.
    A type-$J$ SVN corresponds to a non-empty subset of \acp{VN} that share the same $J$ neighboring \acp{CN}, and is simply referred to as an \ac{SVN}. 
    The total number of SVNs is denoted by~$V$.
    
    With reference to Fig.~\ref{fig:AdjMatrixFairRegular}, \ac{VN}~$1$ and \ac{VN}~$17$ are both connected to \ac{CN}~$1$ (type-$1$) and \ac{CN}~$5$ (type-$2$).
    They therefore belong to the same \ac{SVN}, which consists of all \acp{VN} having the same connections to these two \ac{CN} types.
\end{definition}
\begin{lemma}[Number of SVNs]
\label{lem:num_SVNs}
    For the proposed fair-regular \ac{GLDPC} codes, the number of distinct \acp{SVN} satisfies
    \begin{align}
    V = \min\left\{\left(\frac{M}{J}\right)^J,\, N\right\}.
    \end{align}
\end{lemma}
\begin{proof}
    The result follows from the successive refinements of groups of \acp{VN} according to their connections to the different \ac{CN} types.
    For $x=1$, each \ac{VN} is connected to exactly one of the $M/J$ \acp{CN} belonging to type-$1$.
    Hence, the $N$ \acp{VN} are partitioned into $M/J$ type-$1$ \acp{SVN}.
    When the connection to type $x+1$ is considered, the circulant-lifting pattern of the proposed fair-regular construction distributes the \acp{VN} within each type-$x$ \ac{SVN} according to their connections to the $M/J$ \acp{CN} of type $x+1$.
    Thus, each type-$x$ \ac{SVN} is further refined into up to $M/J$ type-$(x+1)$ \acp{SVN}.
    After considering all $J$ \ac{CN} types, there are $\left(M/J\right)^J$ possible combinations of one \ac{CN} from each type.
    When $\left(M/J\right)^J\leq N$, by construction, every such combination corresponds to an \ac{SVN}, so that $V=\left(M/J\right)^J$.
    On the other hand, if $\left(M/J\right)^J>N$, the construction generates $N$ distinct \ac{CN} neighborhoods before any repetition occurs.
    Since each \ac{SVN} is non-empty, no more than $N$ distinct \acp{SVN} can exist. Hence, $V=N$.
\end{proof}
\begin{remark}
    For fair-regular \ac{GLDPC} codes, all \acp{SVN} have equal cardinality given by
    \begin{align}
    v = \frac{N}{V}.
    \end{align}

    With reference to Fig.~\ref{fig:AdjMatrixFairRegular}, we have $V=16$, yielding $v = N/V = 2$. This means that each \ac{SVN} contains two \acp{VN} sharing the same set of neighboring \acp{CN}.
\end{remark}

\subsection{Stalling Sets}
\label{subsec:StallingSets}

\begin{figure}[t]
    \centering
    \includegraphics[width=1.0\columnwidth]{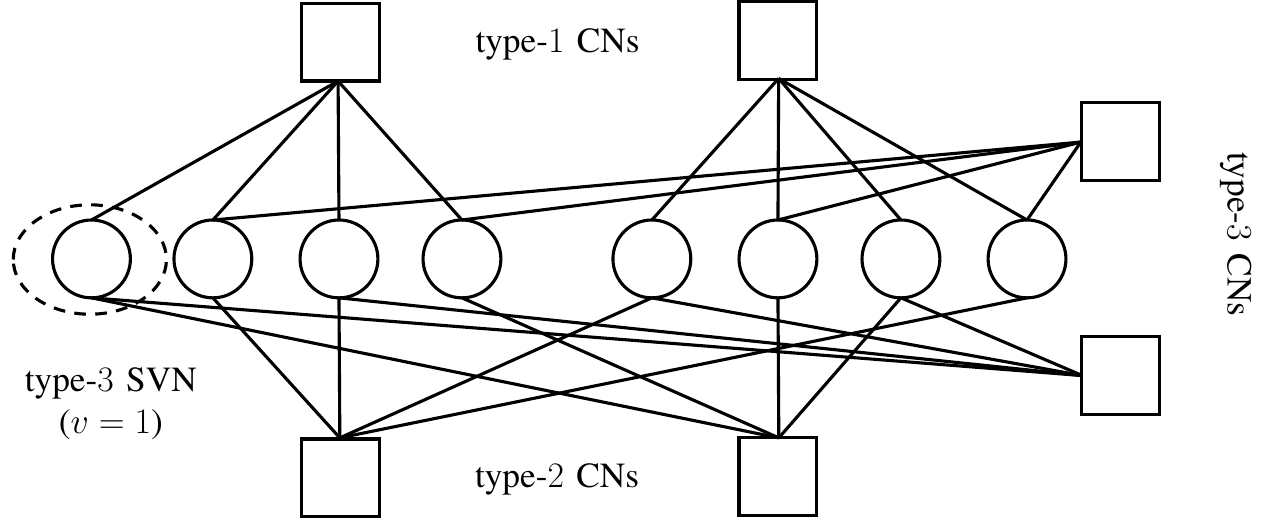}
    \caption{Example of a subgraph involving the \ac{CN}-induced \ac{VN} subset with $|\mathcal{V}| = 8$ \acp{VN} and $J=3$.
    The \ac{SVN} size is $v=1$, and the selected subsets of \acp{CN} are $|\mathcal{S}_1| = 2$, $|\mathcal{S}_2| = 2$, and $|\mathcal{S}_3| = 2$, respectively.}    \label{fig:SharedVNsSubgraph}
\end{figure}

The size of an \ac{SVN} plays a fundamental role in determining the minimum weight of error patterns that can stall the \ac{iBDD} decoder. 
In particular, \acp{SVN} act as elementary building blocks of error configurations, since all \acp{VN} within an \ac{SVN} share the same \ac{CN} neighborhood and are therefore indistinguishable from the decoder perspective.
A smaller value of $v$ prevents errors from concentrating within a single \ac{SVN}, forcing harmful configurations to span multiple \acp{SVN} and increasing the minimum size of the stalling patterns.
Throughout this analysis, we consider an idealized \emph{miscorrection-free} \ac{iBDD} decoder, in which each \ac{CN} either corrects all its erroneous \acp{VN}, if their number does not exceed its correction capability, or otherwise leaves them unchanged, without ever introducing new errors.
\begin{definition}[Stalling set]
\label{def:stalling_set}
A \emph{stalling set} is a non-empty subset of \acp{VN} such that, if all its elements are initially in error, the error pattern remains unchanged across miscorrection-free \ac{iBDD} iterations.
\end{definition}
With reference to Lemma~\ref{lem:num_SVNs}, we conduct the following analysis considering the case $\left(M/J\right)^J\!\leq\! N$, for which the proposed fair-regular construction generates all possible combinations of one \ac{CN} from each \ac{CN} type, and each such combination is associated with an \ac{SVN} of size $v \geq 1$.
Consider non-empty subsets $\{\mathcal{S}_j\}_{j=1}^J$ of \acp{CN}, where $\mathcal{S}_j$ contains \acp{CN} of type $j$, and let $\mathcal{V}$ denote the set of \acp{VN} connected to all selected \acp{CN}.
We refer to $\mathcal{V}$ as the \emph{\ac{CN}-induced \ac{VN} subset}.
Since each combination of one \ac{CN} from every type corresponds to one \ac{SVN}, and each \ac{SVN} contains $v$ \acp{VN}, the cardinality of $\mathcal{V}$ is
\begin{align}
\label{eq:SharedVNS}
|\mathcal{V}| = v \cdot \prod_{j=1}^{J} |\mathcal{S}_j|.
\end{align}
This structured relation between the selected \acp{CN} and the associated
\acp{VN} is what enables the characterization of stalling sets under miscorrection-free \ac{iBDD} developed below.
An example of the corresponding subgraph is illustrated in Fig.~\ref{fig:SharedVNsSubgraph}.
\begin{theorem}[Stalling set condition]
\label{thm:stalling_condition}
Consider a fair-regular \ac{GLDPC} code decoded via miscorrection-free \ac{iBDD}, with component codes of correction capability $t_j$ for type-$j$ \acp{CN}. 
Then, $\mathcal{V}$ contains at least one stalling set if and only if
\begin{align}
\label{eq:StallingSetCondition}
|\mathcal{V}| \ge \max_{j=1,\dots,J} \left\{ (t_j+1)\,|\mathcal{S}_j| \right\}.
\end{align}
\end{theorem}

\begin{proof}
We first prove necessity.
Let $\mathcal{E}\subseteq\mathcal{V}$ be a stalling set, and let $\mathcal{R}_j\subseteq\mathcal{S}_j$ denote the subset of type-$j$ \acp{CN} incident to at least one \ac{VN} in $\mathcal{E}$.
Since $\mathcal{E}$ is a stalling set, every \ac{CN} in $\mathcal{R}_j$ must contain at least $t_j+1$ erroneous \acp{VN}.
Moreover, each erroneous \ac{VN} has exactly one neighbor of type $j$. 
Hence, $|\mathcal{E}| \geq (t_j+1)|\mathcal{R}_j|$, for $j=1,\ldots,J$.
All \acp{VN} in $\mathcal{E}$ belong to the \ac{CN}-induced \ac{VN} subset associated with $\{\mathcal{R}_j\}_{j=1}^{J}$, whose cardinality is $v\prod_{k=1}^{J}|\mathcal{R}_k|$. 
Therefore, $(t_j+1)|\mathcal{R}_j| \leq |\mathcal{E}| \leq v\prod_{k=1}^{J}|\mathcal{R}_k|$,  which implies $t_j+1 \leq v\prod_{\substack{k\neq j}}|\mathcal{R}_k|$.
% $t_j+1 \leq v\prod_{\substack{k=1\\k\neq j}}^{J} |\mathcal{R}_k|$.
%
Since $\mathcal{R}_k\subseteq\mathcal{S}_k$, we have $t_j+1 \leq 
v\prod_{\substack{k\neq j}} 
% v\prod_{\substack{k=1\\k\neq j}}^{J} 
|\mathcal{S}_k|$.
Multiplying by $|\mathcal{S}_j|$ and using \eqref{eq:SharedVNS}, we obtain $|\mathcal{V}| \geq (t_j+1)|\mathcal{S}_j|$, for $j=1,\ldots,J$, which is equivalent to~\eqref{eq:StallingSetCondition}.

Conversely, suppose that~\eqref{eq:StallingSetCondition} holds.
Then, for every type $j$, each \ac{CN} in $\mathcal{S}_j$ is connected to ${|\mathcal{V}|}/{|\mathcal{S}_j|} \geq t_j+1$  \acp{VN} in $\mathcal{V}$.
Therefore, if all \acp{VN} in $\mathcal{V}$ are initially in error, every selected type-$j$ \ac{CN} observes more errors than its correction capability and, under miscorrection-free \ac{iBDD}, performs no update.
Furthermore, \acp{CN} outside $\{\mathcal{S}_j\}_{j=1}^{J}$ are connected to no \acp{VN} in $\mathcal{V}$.
Hence, the error pattern remains unchanged across iterations, and $\mathcal{V}$ itself is a stalling set.
\end{proof}

It is worth emphasizing that Theorem~\ref{thm:stalling_condition} characterizes the existence of a stalling set within the considered \ac{CN}-induced \ac{VN} subset $\mathcal{V}$ under miscorrection-free \ac{iBDD}; it does not characterize the minimum distance of the overall \ac{GLDPC} code.
Determining the minimum distance of a GLDPC code for a given graph and set of component codes is, in general, a challenging problem~\cite{smarandache2025generalizedquasicyclicldpccodes,smarandache2025generalizedquasicyclicldpccodes_ISIT}.
\begin{lemma}[Balanced selection of variable nodes]
\label{lemma:balanced_VNs}
Consider a \ac{CN}-induced \ac{VN} subset $\mathcal{V}$
associated with the \ac{CN} subsets
$\{\mathcal{S}_j\}_{j=1}^{J}$.
For any integer $0\leq e\leq|\mathcal{V}|$, there exists a
selection of $e$ \acp{VN} from $\mathcal{V}$ such that, for
every type $j$, the number of selected \acp{VN} connected to
each \ac{CN} in $\mathcal{S}_j$ is either
\begin{align}
\left\lfloor
\frac{e}{|\mathcal{S}_j|}
\right\rfloor
\quad\text{or}\quad
\left\lceil
\frac{e}{|\mathcal{S}_j|}
\right\rceil.
\end{align}
\end{lemma}
\begin{proof}
See Appendix~A.
\end{proof}
For a given selection $\{\mathcal{S}_j\}_{j=1}^{J}$, any stalling set supported by these \acp{CN} must contain at least $\max_j\left\{(t_j+1)|\mathcal{S}_j|\right\}$
erroneous \acp{VN}. By Lemma~\ref{lemma:balanced_VNs}, this minimum weight is achievable whenever the corresponding \ac{CN}-induced \ac{VN} subset is sufficiently large, namely when \eqref{eq:StallingSetCondition} holds.
Therefore, the minimum stalling set weight is obtained by minimizing the above quantity over all feasible \ac{CN} selections.

We therefore define $\{\mathcal{S}_j^{\min}\}_{j=1}^{J}$ as a feasible \ac{CN} selection minimizing the required number of erroneous \acp{VN}, i.e., 
\begin{align}
\left\{ |\mathcal{S}_j^{\min}| \right\}_{j=1}^{J} 
&\in \underset{\{|\mathcal{S}_j|\}}{\arg\min} \max_j \left\{ (t_j+1)|\mathcal{S}_j| \right\} \notag\\ 
&\text{s.t.}\quad v\prod_{j=1}^{J}|\mathcal{S}_j| \geq \max_j \left\{ (t_j+1)|\mathcal{S}_j| \right\}.
\label{eq:min_candidate_CN_selection}
\end{align}
In the following, in case multiple selections $\{|\mathcal{S}_j|\}_{j=1}^{J}$ achieve the same minimum value, we select the one inducing the smallest \ac{CN}-induced \ac{VN} subset $\mathcal{V}$, i.e., the one minimizing $|\mathcal{V}|$. After this filtering, one of the remaining selections is chosen arbitrarily.

For example, consider the $J=3$ case with $v=1$ and $t_j=3$ for every $j$.
Starting from the smallest possible selection, $|\mathcal{S}_1|=|\mathcal{S}_2|=|\mathcal{S}_3|=1$, the \ac{CN}-induced \ac{VN} subset contains $|\mathcal{V}|=1$ \ac{VN}, which is insufficient to satisfy the feasibility condition in~\eqref{eq:min_candidate_CN_selection}.
Increasing the number of selected \acp{CN} of one type to $|\mathcal{S}_1|=2$ gives $|\mathcal{V}|=2$, and selecting two \acp{CN} from two types, i.e., $|\mathcal{S}_1|=|\mathcal{S}_2|=2$ and $|\mathcal{S}_3|=1$, gives $|\mathcal{V}|=4$.
Both selections remain infeasible.
Finally, selecting two \acp{CN} from each type, as illustrated in Fig.~\ref{fig:SharedVNsSubgraph}, gives $|\mathcal{V}| = 1\cdot2\cdot2\cdot2 = 8$.
Each selected \ac{CN} is then connected to $|\mathcal{V}|/|\mathcal{S}_j|=4=t_j+1$ \acp{VN}.
Hence, the selection $|\mathcal{S}_1^{\min}|=|\mathcal{S}_2^{\min}|=|\mathcal{S}_3^{\min}|=2$ is feasible and yields $\max_j\{(t_j+1)|\mathcal{S}_j^{\min}|\} = 8$.

\begin{theorem}[Minimum stalling set weight]
\label{thm:minimum_stalling_set}
Under miscorrection-free \ac{iBDD}, the minimum stalling set weight is
\begin{align}
\label{eq:min_stalling_weight}
w_{\min} =  \max_j \left\{ (t_j+1) |\mathcal{S}_j^{\min}| \right\}\,.
\end{align}
\end{theorem}
\begin{proof}
Let $\mathcal{E}$ be an arbitrary stalling set of weight $w=|\mathcal{E}|$, and let $\mathcal{S}_j$ denote the set of type-$j$ \acp{CN} incident to $\mathcal{E}$.
Since every \ac{VN} has exactly one neighbor of each type, the $w$ erroneous \acp{VN} contribute exactly $w$ erroneous edges toward the type-$j$ \acp{CN}.
Every \ac{CN} in $\mathcal{S}_j$ must contain at least $t_j+1$ errors, and therefore $w \geq (t_j+1)|\mathcal{S}_j|$, for $j=1,\ldots,J$.
Hence, $w \geq \max_j \left\{ (t_j+1)|\mathcal{S}_j| \right\}$.
Moreover, $\mathcal{E}$ is contained in the \ac{CN}-induced \ac{VN} subset associated with $\{\mathcal{S}_j\}_{j=1}^{J}$, so $|\mathcal{V}| \geq |\mathcal{E}| \geq \max_j \left\{ (t_j+1)|\mathcal{S}_j| \right\}$.
Thus, $\{\mathcal{S}_j\}_{j=1}^{J}$ is a feasible selection
in~\eqref{eq:min_candidate_CN_selection}, which implies $w \geq \max_j \left\{ (t_j+1)|\mathcal{S}_j^{\min}| \right\}$.
To show achievability, define $e^\star = \max_j \left\{ (t_j+1)|\mathcal{S}_j^{\min}| \right\}$.
By feasibility, $e^\star \leq v\prod_{j=1}^{J} |\mathcal{S}_j^{\min}|$.
Lemma~\ref{lemma:balanced_VNs} therefore guarantees the existence of a subset of $e^\star$ \acp{VN} such that every type-$j$ \ac{CN} in $\mathcal{S}_j^{\min}$ is connected to at least $\left\lfloor
e^\star/|\mathcal{S}_j^{\min}|
\right\rfloor
\geq
t_j+1$ erroneous \acp{VN}.
Hence, this subset is a stalling set and the lower bound on $w$ is achieved.
\end{proof}
\begin{remark}
For $v>\max_j(t_j)$, the minimizing selection is $|\mathcal{S}_j^{\min}|=1$ for every $j$, corresponding to a single \ac{SVN}. Hence,
\begin{align}
w_{\min}=\max_j(t_j+1).
\end{align}
For $v\leq\max_j(t_j)$, a single \ac{SVN} is insufficient, and the minimizing selection involves multiple \acp{SVN}.
\end{remark}

%%%%%%%%%%%%%%%%%%%%%%%%%%%%%%%%%%%%%%%%%%%%%%%%%%%%%%
%%%%%%%%%%%%%%%%%%%%%%%%%%%%%%%%%%%%%%%%%%%%%%%%%%%%%%

%%%%%%%%%%%%%%%%%%%%%%%%%%%%%%%%%%%%%%%%%%%%%%%%%%%%%%
%%%%%%%%%%%%%%%%%%%%%%%%%%%%%%%%%%%%%%%%%%%%%%%%%%%%%%

%%%%%%%%%%%%%%%%%%%%%%%%%%%%%%%%%%%%%%%%%%%%%%%%%%%%%%
%%%%%%%%%%%%%%%%%%%%%%%%%%%%%%%%%%%%%%%%%%%%%%%%%%%%%%
\section{Finite-Length Error Floor Analysis}
\label{sec:ErrorFloorAnalysis}
 
Throughout this section, we retain the miscorrection-free \ac{iBDD} model introduced in Section~\ref{subsec:StallingSets}. 
The resulting expressions are rigorous bounds for this idealized decoder. 
Their applicability as error floor estimates for conventional \ac{iBDD} is assessed numerically in Section \ref{sec:NumericalResults}.
In the following, we characterize the minimum-weight stalling sets that can form in the Tanner graph, showing that their structure depends jointly on the component-code correction capabilities $t_j$, the number of \acp{CN} of each type $j$ required to induce them, and the \ac{SVN} size $v$, which together give rise to two distinct operating regimes.
The analysis can be extended to other \ac{GLDPC} or product-like constructions whose \ac{VN}-\ac{CN} connectivity admits an analogous partition of \acp{VN} into \acp{SVN}.
In such cases, the stalling set condition of Theorem~\ref{thm:stalling_condition} and the minimum stalling set weight characterization of Theorem~\ref{thm:minimum_stalling_set} can be applied after appropriately characterizing the corresponding SVN and \ac{CN}-induced \ac{VN} subset structures.

Let $w_\mathrm{min}$ be the minimum stalling set weight, and let $A_{w_{\min}}$ denote the number of minimum-weight stalling set configurations.
According to Definition~\ref{def:stalling_set}, each such configuration necessarily results in decoding failure under miscorrection-free \ac{iBDD}, since none of the erroneous \acp{VN} can be corrected.
The \ac{CER} can thus be lower bounded as
\begin{align}
\label{eq:CER_LB}
\text{CER}_\text{MF}
\ge
A_{w_{\min}}\,p^{w_{\min}}(1-p)^{N-w_{\min}},
\end{align}
where $p$ denotes the equivalent \ac{HD} channel error probability seen by the \ac{iBDD} decoder.\footnote{Specifically, the considered transmission model is a \ac{bi-AWGN} channel with \ac{BPSK} modulation, followed by \ac{HD} at the receiver.
This induces an equivalent \ac{BSC} with crossover probability $p = (1/2) \text{erfc}\sqrt{R E_\mathrm{b}/N_0}$.}
Accordingly, the \ac{BER} satisfies
\begin{align}
\label{eq:BER_LB}
\text{BER}_\text{MF}
\ge
\frac{w_{\min}}{N}\,
A_{w_{\min}}\,p^{w_{\min}}(1-p)^{N-w_{\min}}.
\end{align}
In the low error rate regime (small $p$), these minimum-weight configurations dominate the error floor behavior.
\begin{figure}[t]
    \centering
    \includegraphics[width=1.0\columnwidth]{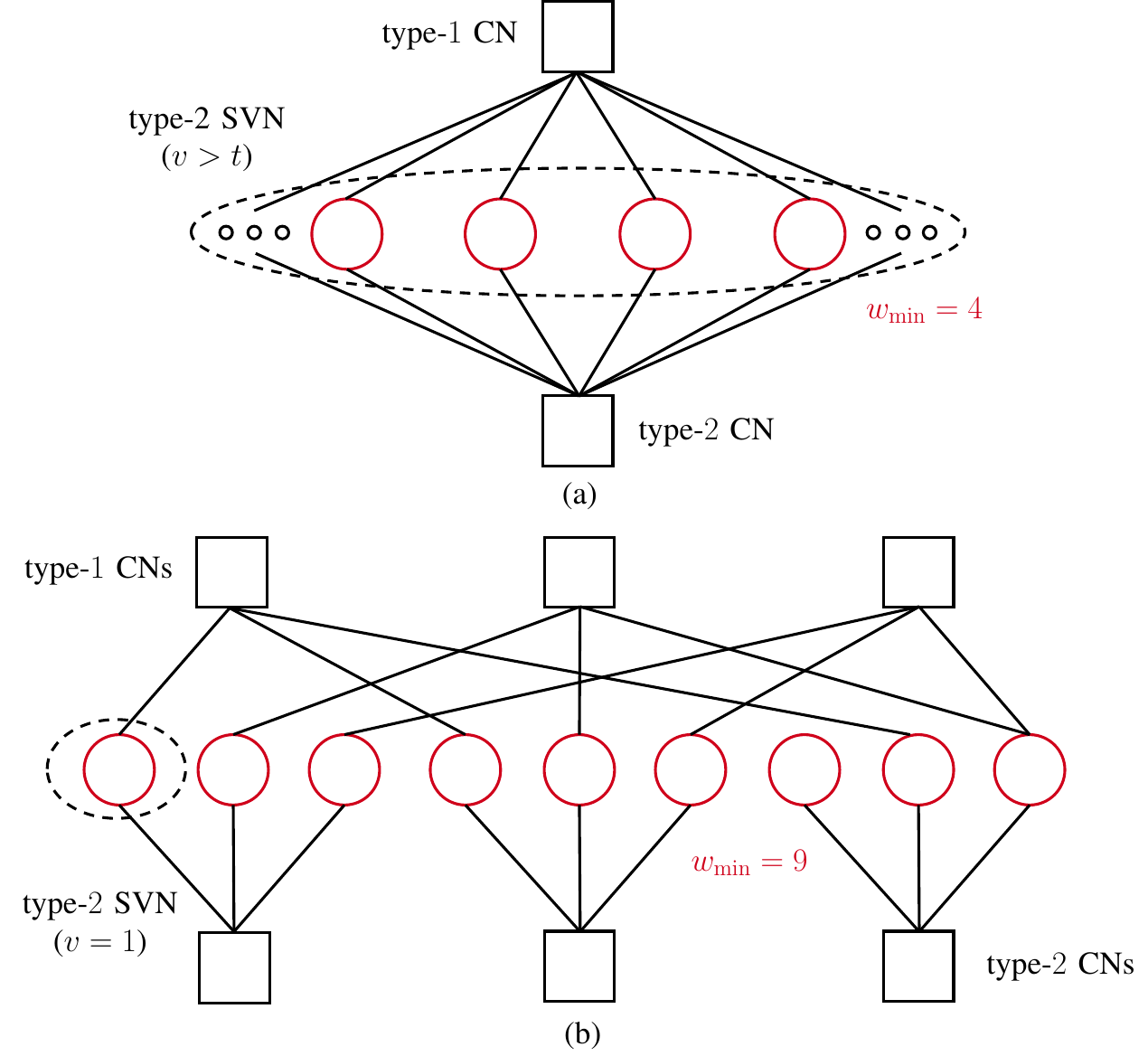}
    \caption{Examples of minimum stalling sets for miscorrection-free \ac{iBDD} in \ac{GLDPC} codes with $J=2$. (a) Stalling set in the regime $v > \max_j (t_j)$ with $t_j=3$ for all $j$, where the minimum stalling pattern has weight $w_\mathrm{min}=t+1=4$ and is contained within a single \ac{SVN}, leading to decoder stalling.
    (b) Stalling set in the regime $v \le \max_j (t_j)$ with $t_j=2$ for all $j$, where the smallest stalling configuration has weight $w_\mathrm{min}=(t+1)|\mathcal{S}_j^{\min}|=9$ and spans $|\mathcal{S}_j^{\min}|=3$ \acp{CN} of each type.}
    \label{fig:StallingSets_J2}
\end{figure}
\begin{theorem}[Error floor in the high-$v$ regime]
\label{thm:high_v_error_floor}
Consider fair-regular \ac{GLDPC} codes under miscorrection-free \ac{iBDD}, with \ac{SVN} size satisfying $v > \max_j(t_j)$.
In this regime, we have
\begin{align}
\text{\emph{CER}}_{\text{\emph{MF}}}
\ge
V\binom{v}{w_{\min}}
p^{w_{\min}}(1-p)^{N-w_{\min}}\,.
\end{align}
where $w_{\min}=\max_j(t_j+1)$.
\end{theorem}
\begin{proof}
    Since $v > \max_j(t_j)$, Theorem~\ref{thm:minimum_stalling_set} with the minimizing \ac{CN} selection provided in~\eqref{eq:min_candidate_CN_selection} gives $w_{\min} = \max_j(t_j+1)$.
    Each \ac{SVN}, containing $v$ \acp{VN}, contributes at least $\binom{v}{w_{\min}}$ harmful patterns of minimum weight.
    Since there are $V$ \acp{SVN}, the multiplicity of the dominant configurations satisfies $A_{w_{\min}} \ge V\binom{v}{w_{\min}}$.
\end{proof}
An example of such a stalling set configuration is illustrated in Fig.~\ref{fig:StallingSets_J2}(a) for the case $t_j=3$ for all $j$.
For single-class component codes, i.e., $t_j=t$ for all $j$, the result reduces to $w_{\min} = t+1$.

\begin{theorem}[Error floor in the low-$v$ regime]
\label{thm:low_v_error_floor}
Consider fair-regular \ac{GLDPC} codes under miscorrection-free \ac{iBDD}, with \ac{SVN} size satisfying $v \le \max_j(t_j)$.
In this regime, we have
\begin{align}
\text{\emph{CER}}_{\text{\emph{MF}}}
\ge C_{w_{\min}}
\left(
\prod_{j=1}^{J}
\binom{M/J}{|\mathcal{S}_j^{\min}|}
\right)
p^{w_{\min}}
(1-p)^{N-w_{\min}}\,.
\end{align}
where $C_{w_{\min}}$ is the number of distinct minimum-weight stalling sets contained within the \ac{CN}-induced \ac{VN} subset associated with the minimizing selection $\{\mathcal{S}_j^{\min}\}_{j=1}^{J}$, and $w_{\min}=\max_j\left\{(t_j+1)|\mathcal{S}_j^{\min}|\right\}$.
\end{theorem}
\begin{proof}
Since $v \leq \max_j(t_j)$, a single \ac{SVN} is insufficient to form a minimum-weight stalling set, and the corresponding configuration must span multiple \acp{SVN}.
By Theorem~\ref{thm:minimum_stalling_set}, the minimum stalling set weight is $w_{\min}=\max_j \left\{(t_j+1)|\mathcal{S}_j^{\min}| \right\}$, where $\{\mathcal{S}_j^{\min}\}_{j=1}^{J}$ is the minimizing \ac{CN} selection given in~\eqref{eq:min_candidate_CN_selection}.
A lower bound on the multiplicity of dominant configurations is obtained by counting admissible selections of \acp{CN} and the number of minimum-weight stalling sets inside them, yielding $A_{w_{\min}} \ge
C_{w_{\min}} \left(\prod_{j=1}^{J} \binom{M/J}{|\mathcal{S}_j^{\min}|}\right)$.
\end{proof}
\begin{remark}
The stalling set definition considered in this work focuses on initial error patterns for which none of the involved \acp{CN} can successfully correct the erroneous \acp{VN} under miscorrection-free \ac{iBDD}, which provides a direct and analytically tractable characterization of error floor events.
In practice, however, additional harmful configurations may emerge during iterative decoding as a result of miscorrections produced by previously processed \acp{CN}, as observed in iterative \ac{HD} decoding of product and staircase-like codes.
Accounting for such effects analytically is difficult, and they are typically addressed relying on empirical or simulation-based characterizations rather than closed-form analysis~\cite{Smith2012:StaircaseCodes}.

For example, in the low-$v$ regime and for heterogeneous designs, a \ac{CN} may observe more than $t_j$ errors within an otherwise minimum-weight error configuration and may consequently produce a miscorrection rather than fail to correct the errors, potentially introducing new, dynamically generated stalling configurations.
Nonetheless, numerical results show that the proposed fair-regular GLDPC codes under conventional iBDD tightly approach the miscorrection-free lower bounds in the region of interest, indicating that problematic miscorrections occur with negligible probability.
Suitable scheduling strategies and miscorrection-detection mechanisms, such as syndrome-based processing or the error-detection capabilities inherently available in \acp{eBCH}, may help further mitigate the formation and propagation of such harmful configurations.
\end{remark}
For specific parameter choices, the multiplicity can be further refined through exact enumeration $C_{w_{\min}}$ of the valid minimum-weight stalling sets within the induced subgraph.

\paragraph*{Example 1 ($J=2,\, t_j=2$ for all $j$)}
The minimum stalling set weight is $w_{\min}=9$, obtained with $|\mathcal{S}_j^{\min}|=3$ for all $j$. 
The corresponding stalling configuration is illustrated in Fig.~\ref{fig:StallingSets_J2}(b). 
In this case, a unique error configuration exists within the induced set ($C_{w_{\min}} = 1$), yielding
\begin{align}
\text{{CER}}_{\text{{MF}}} \ge \left[\binom{M/2}{3}\right]^2 p^9 (1-p)^{N-9}.
\end{align}

\begin{figure}[t]
    \centering
        % \resizebox{\linewidth}{!}{\input{Figures/StallingSets_J3}}
        \includegraphics[width=1.0\columnwidth]{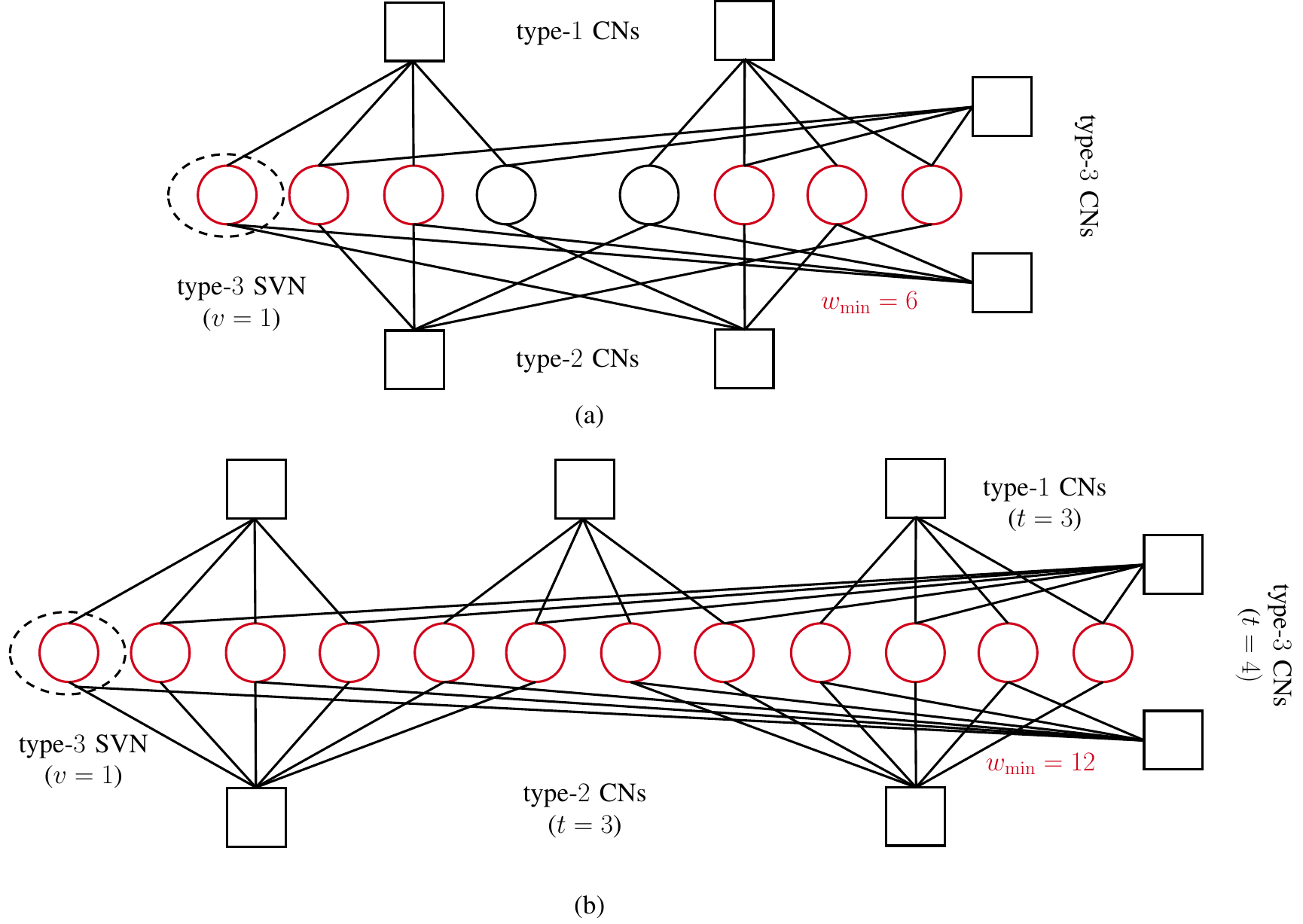}
    \caption{Examples of minimum stalling sets for miscorrection-free \ac{iBDD} in \ac{GLDPC} codes with $J=3$.
    (a) Stalling set in the regime $v \le \max_j (t_j)$ with $t_j=2$ for all $j$, where the minimum stalling pattern has weight $w_\mathrm{min}=6$ and involves $|\mathcal{S}_j^{\min}|=2$ \acp{CN} for each type $j$.
    (b) Stalling set in the regime $v \le \max_j (t_j)$ for the heterogeneous case with mixed component code capabilities ($t_j \in \{3,4\}$), where the minimum stalling pattern has weight $w_\mathrm{min}=12$ and involves $|\mathcal{S}_1^{\min}|=3$ \acp{CN} with $t_1=3$, $|\mathcal{S}_2^{\min}|=2$ \acp{CN} with $t_2=3$, and $|\mathcal{S}_3^{\min}|=2$ \acp{CN} with $t_3=4$.}
    \label{fig:StallingSets_J3}
\end{figure}

\paragraph*{Example 2 ($J=3,\, t_j=2$ for all $j$)}
The minimum stalling set weight is $w_{\min}=6$, obtained with $|\mathcal{S}_j^{\min}|=2$ for all $j$. 
In this case, $C_{w_{\min}} = 4$ valid error configurations exist within the induced set, yielding
\begin{align}
\text{{CER}}_{\text{{MF}}} \ge 4 \cdot \left[\binom{M/3}{2}\right]^3 p^6 (1-p)^{N-6}.
\end{align}
A possible minimum-weight error pattern for the considered scenario is illustrated in Fig.~\ref{fig:StallingSets_J3}(a).

\paragraph*{Example 3 ($J=3,\, t_j=3$ for all $j$)}
The minimum stalling set weight is $w_{\min}=8$, again with $|\mathcal{S}_j^{\min}|=2$ for all $j$. 
In this scenario, only one configuration is admissible ($C_{w_{\min}}=1$), yielding
\begin{align}
\text{{CER}}_{\text{{MF}}} \ge \left[\binom{M/3}{2}\right]^3 p^8 (1-p)^{N-8}.
\end{align}

\paragraph*{Example 4 ($J=3,\, t_j=4$ for all $j$)}
The minimum configuration satisfying~\eqref{eq:min_candidate_CN_selection} involves an asymmetric selection of \acp{CN}, i.e., $|\mathcal{S}_j^{\min}| = \{3,3,2\}$ (up to permutations), leading to a minimum stalling set weight $w_{\min}=15$ and $C_{w_{\min}} =6$ admissible configurations. 
This yields
\begin{align}
\text{{CER}}_{\text{{MF}}} \ge 6 \cdot \left[\binom{M/3}{3}\right]^2 \cdot \binom{M/3}{2} \, p^{15} (1-p)^{N-15}.
\end{align}

\paragraph*{Example 5 ($J=3$, $(t_1,t_2,t_3) = (3,3,4)$)}
The minimum configuration satisfying~\eqref{eq:min_candidate_CN_selection} is obtained with $|\mathcal{S}_j^{\min}| = \{3,2,2\}$ (only $2$ type-$3$ \acp{CN} are involved), leading to a minimum stalling set weight $w_{\min}=12$ and a single admissible configuration ($C_{w_{\min}} = 1$). 
The corresponding multiplicity is dominated by the number of \ac{CN} selections, yielding
\begin{align}
\text{{CER}}_{\text{{MF}}} \ge \left[\binom{M/3}{2}\right]^2 \cdot \binom{M/3}{3} \, p^{12} (1-p)^{N-12}.
\end{align}
A possible minimum-weight error pattern for the considered scenario is illustrated in Fig.~\ref{fig:StallingSets_J3}(b).

%

%%%%%%%%%%%%%%%%%%%%%%%%%%%%%%%%%%%%%%%%%%%%%%%%%%%%%%
%%%%%%%%%%%%%%%%%%%%%%%%%%%%%%%%%%%%%%%%%%%%%%%%%%%%%%

\section{Approximate Waterfall Analysis}
\label{sec:AsymptoticWaterfallAnalysis}

We develop an analytical surrogate to evaluate the waterfall performance of the proposed fair-regular \ac{GLDPC} codes under conventional \ac{iBDD}.
The analysis is based on an independence assumption at the inputs to the component decoders, whereby the error states of the \acp{VN} entering each component decoding are modeled as statistically independent Bernoulli random variables.
Accordingly, the dependencies that may arise from the structured graph and from the \ac{VN} updates produced by earlier component decoding operations are neglected.
This formulation provides a computationally efficient prediction of the waterfall behavior and offers insights into the achievable \ac{NCG} of the considered \ac{GLDPC} constructions.
The fair-regular deterministic structure plays a complementary role in the proposed design methodology, as the controlled \ac{VN}-\ac{CN} connectivity described through the \ac{SVN} structure enables the dominant finite-length stalling sets to be identified and enumerated, supporting error floor prediction.
Thus, the deterministic construction enables the finite-length error floor analysis, while the independence-based recursion described in this section provides an efficient analytical surrogate for waterfall prediction.

Let $p_\ell$ denote the probability that a randomly selected \ac{VN} is in error at iteration $\ell$.
Each \ac{VN} is connected to exactly one \ac{CN} per type.
Under the aforementioned independence assumption, and considering the homogeneous case in which all \ac{CN} types employ component codes with the same decoding rule, the evolution of the bit error probability can be approximated as
\begin{align}
\label{eq:pEvolution}
p_{\ell+1} = (\underbrace{F \circ F \circ \cdots \circ F}_{J \text{ times}}) (p_\ell),
\end{align}
where $F(\cdot)$ denotes the update rule induced by a \ac{CN} type.\footnote{This formulation can be naturally extended to heterogeneous component codes by introducing type-dependent update functions $F_j(\cdot)$.}
This recursion provides an approximate evolution of the bit error probability for the finite-length fair-regular \ac{GLDPC} code family.
It should therefore be interpreted as an analytical surrogate for the deterministic construction, rather than as an exact or rigorous asymptotic characterization of its decoding behavior.
Let $e_i \in \{0,1\}$ denote the channel error at position $i$, and let $p = \mathbb{P}(e_i=1)$ denote the input error probability.
Due to symmetry of linear codes over the \ac{BSC}, we assume without loss of generality that the transmitted codeword is the all-zero codeword, i.e., $\V{c} = \V{0}$.
The bit error probability after \ac{BDD} is
\begin{align}
\label{eq:F_pUpdate}
F(p) \! = \! p \, \mathbb{P}(\hat{x}_i=1 \! \mid \! e_i=1) \! + \! (1\!-\!p) \, \mathbb{P}(\hat{x}_i=1 \! \mid \! e_i=0).
\end{align}
Let $\V{e} \in \{0,1\}^{n}$ be the channel error vector, whose entries are \ac{i.i.d.} Bernoulli random variables with parameter $p$, and let $w_\mathrm{H}(\V{e}) = e$ denote its Hamming weight.
After \ac{BDD}, the decoder may output a codeword $\hat{\V{c}}$.
We define three mutually exclusive events: $(i)$ correct decoding $\mathcal{C}$ ($\hat{\V{c}} = \V{c} = \V{0}$); $(ii)$ decoding failure $\mathcal{F}$ (no codeword selected, error detection); $(iii)$ miscorrection $\mathcal{M}$ ($\hat{\V{c}} = \tilde{\V{c}} \neq \V{0}$).

We first consider the case where the bit in position $i$ is initially in error, i.e., $e_i = 1$.
In this case, the bit remains in error after \ac{BDD} if either decoding fails or a miscorrection does not correct it:
\begingroup
\allowdisplaybreaks
\begin{align}
\label{eq:p_HatX1_GivenE1}
&\mathbb{P}(\hat{x}_i=1 \mid e_i=1)  \notag \\
&= \mathbb{P}(\mathcal{F} \mid e_i=1) + \mathbb{P}(\hat{x}_i=1 \mid e_i=1, \mathcal{M}) \cdot \mathbb{P}(\mathcal{M} \mid e_i=1) \notag \\
&\leq \mathbb{P}(\mathcal{F} \mid e_i=1) + \mathbb{P}(\mathcal{M} \mid e_i=1) \notag \\
&= \mathbb{P}(w_\mathrm{H}(\V{e}) \ge t+1 \mid e_i = 1) \notag\\
&= \sum_{e=t}^{n-1} \binom{n-1}{e} p^e (1-p)^{n-1-e}.
\end{align}
\endgroup
The bound is particularly tight for large enough $n$, as the probability that a miscorrection corrects the bit is very low.

We now consider the case where the bit is initially correct, i.e., $e_i=0$. 
In this case, errors can only arise due to miscorrections, so we have
\begingroup
\allowdisplaybreaks
\begin{align}
\label{eq:p_HatX1_GivenE0}
    &\mathbb{P}(\hat{x}_i\!=\!1 \! \mid \! e_i\!=\!0) \! = 
    \mathbb{P}(\hat{x}_i\!=\!1 \! \mid \! e_i\!=\!0, \mathcal{M}) \! \cdot \! \mathbb{P}(\mathcal{M} \! \mid \! e_i\!=\!0) .
\end{align}
\endgroup
We first compute the conditional probability that a bit is flipped given that a miscorrection occurs, $\mathbb{P}(\hat{x}_i\!=\!1 \! \mid \! e_i\!=\!0, \mathcal{M})$.
Let $\mathcal{M}_d$ denote the event that \ac{BDD} miscorrects to a codeword of Hamming weight $d$.
By the law of total probability we obtain
\begingroup
\allowdisplaybreaks
\begin{align}
\label{eq:p_HatX1_GivenE0_Miscor}
&\mathbb{P}(\hat{x}_i=1 \mid e_i=0, \mathcal{M}) \notag \\
&=\sum_{d=d_{\min}}^{n} \!\!\! \mathbb{P}(\hat{x}_i=1 \mid e_i = 0, \mathcal{M}_d) \cdot \mathbb{P}(\mathcal{M}_d \mid e_i=0, \mathcal{M}).
\end{align}
\endgroup
Given a miscorrection to a codeword of weight $d$, the probability that an initially correct bit in position $i$ belongs to its support is upper bounded by
\begin{align}
\label{eq:p_HatX1_GivenE0_Mis_d}
\mathbb{P}(\hat{x}_i=1 \mid e_i=0, \mathcal{M}_d) \le \frac{t}{n - (d-t)}\,.
\end{align}
Moreover, since $\mathcal{M}_d \subseteq \mathcal{M}$, by the definition of conditional probability, we have
\begingroup
\allowdisplaybreaks
\begin{align}
\label{eq:p_Mis_d_GivenE0_Mis}
&\mathbb{P}(\mathcal{M}_d \mid e_i=0, \mathcal{M}) \notag \\
&= \frac{\mathbb{P}(\mathcal{M}_d \, \cap \, \mathcal{M}\mid e_i=0)}
{\mathbb{P}(\mathcal{M} \mid e_i=0)} = \frac{\mathbb{P}(\mathcal{M}_d \mid e_i=0)}
{\mathbb{P}(\mathcal{M} \mid e_i=0)}\,.
\end{align}
\endgroup
A miscorrection under \ac{BDD} occurs when the received sequence lies within the decoding region of an incorrect codeword.
Hence, using the union bound, we can write
\begingroup
\allowdisplaybreaks
\begin{align}
\label{eq:p_Mis_d_GivenE0_only} 
\mathbb{P}(\mathcal{M}_d \mid e_i=0) &= \mathbb{P} \left( \cup_{\tilde{\V{c}}} \,\, \mathcal{M}_d(\tilde{\V{c}}) \mid e_i=0 \right) \notag \\
&\le \sum_{\tilde{\V{c}}}
\mathbb{P}(\mathcal{M}_d(\tilde{\V{c}}) \mid e_i=0) \notag \\
&= A_d \, \mathbb{P}(\mathcal{M}_d(\tilde{\V{c}}) \mid e_i=0).
\end{align}
\endgroup
In~\eqref{eq:p_Mis_d_GivenE0_only}, $A_d$ denotes the multiplicity of codewords of weight $d$, and $\mathcal{M}_d(\tilde{\V{c}})$ denotes the event that \ac{BDD} miscorrects to a codeword $\tilde{\V{c}}$ of Hamming weight $d$ such that $w_\mathrm{H} (\tilde{\V{c}} + \V{e}) < w_\mathrm{H} ({\V{e}})$.
Substituting~\eqref{eq:p_Mis_d_GivenE0_only} in~\eqref{eq:p_Mis_d_GivenE0_Mis},~\eqref{eq:p_Mis_d_GivenE0_Mis} and~\eqref{eq:p_HatX1_GivenE0_Mis_d} in~\eqref{eq:p_HatX1_GivenE0_Miscor}, and~\eqref{eq:p_HatX1_GivenE0_Miscor} in~\eqref{eq:p_HatX1_GivenE0}, the    $\mathbb{P}(\mathcal{M} \mid e_i=0)$ cancels out, yielding
\begin{align}
\label{eq:p_HatX1_GivenE0_Bound}
\mathbb{P}(\hat{x}_i\!=\!1 \! \mid \! e_i\!=\!0) \! \le \!\!\!\! \sum_{d=d_{\min}}^{n} \!\!\!\! \frac{t}{n \! - \! (d \!- \! t)} A_d \, \mathbb{P}(\mathcal{M}_d(\tilde{\V{c}}) \! \mid \! e_i\!=\!0).
\end{align}

We now derive the probability $\mathbb{P}(\mathcal{M}_d(\tilde{\V{c}}) \mid e_i=0)$.
Let $e$ denote the number of errors among the $n-1$ remaining positions excluding $i$, and let $q$ denote the number of these error positions that also belong to the support of the competing codeword $\tilde{\V{c}}$.
Since BDD can decode to $\tilde{\V{c}}$ only if the received sequence lies within Hamming distance $t$ of $\tilde{\V{c}}$, a necessary condition for the miscorrection is
\begingroup
\allowdisplaybreaks
\begin{align}
\label{eq:Misc_condition}
w_\mathrm{H}(\tilde{\V{c}} + \V{e}) = d + e - 2q \le t\,.
\end{align}
\endgroup
Moreover, since $q\leq e$ and $q\leq d$, the above condition implies $e \ge d-t=e_{\min}$ and $e\le d+t=e_{\max}$.
Thus, only error patterns with $e_{\min}\leq e\leq e_{\max}$ need to be considered.
Due to the conditioning $e_i=0$, we distinguish two cases depending on whether or not the competing codeword has a `$1$' in position~$i$.
In particular, by applying the law of total probability, we upper bound the resulting expression as
\begingroup
\allowdisplaybreaks
\begin{align}
\label{eq:p_Mis_d_GivenE0}
\mathbb{P}(\mathcal{M}_d(\tilde{\V{c}}) \mid e_i=0) \le \max \{&\mathbb{P}(\mathcal{M}_d(\tilde{\V{c}}) \! \mid \! e_i\!=\!0, \tilde{c}_i\!=\!1), \notag \\
&\mathbb{P}(\mathcal{M}_d(\tilde{\V{c}}) \! \mid \! e_i\!=\!0, \tilde{c}_i\!=\!0)\}.
\end{align}
\endgroup
Conditioning on $\tilde{c}_i=1$, we obtain
\begingroup
\allowdisplaybreaks
\begin{align}
\label{eq:p_Mis_d_GivenE0_C1}
&\mathbb{P}(\mathcal{M}_d(\tilde{\V{c}}) \mid e_i = 0, \tilde{c}_i = 1) \notag \\
&= \sum_{e=e_\mathrm{min}}^{e_{\max}} \mathbb{P}(\mathcal{M}_d(\tilde{\V{c}}) \mid e_i=0, \tilde{c}_i = 1, w_\mathrm{H}({\V{e}}) = e) \notag \\
&\quad\times \mathbb{P}(w_\mathrm{H}({\V{e}}) = e \mid e_i = 0) \notag \\
&= \sum_{e=e_\mathrm{min}}^{e_{\max}} \sum_{q =q_\mathrm{min}^{(1)}}^{q_\mathrm{max}^{(1)}} \binom{d-1}{q} \binom{n-d}{e-q} p^e (1-p)^{n-1-e}\,,
\end{align}
\endgroup
where $q_\mathrm{min}^{(1)} = \max\{\left\lfloor d/2\right\rfloor+1, \left\lceil (d+e-t)/2\rceil, e-(n-d)\right\}$ and $q_\mathrm{max}^{(1)}= \min\{d-1,e\}$.
On the other hand, conditioning on $\tilde{c}_i=0$, we have
\begingroup
\allowdisplaybreaks
\begin{align}
\label{eq:p_Mis_d_GivenE0_C0}
&\mathbb{P}(\mathcal{M}_d(\tilde{\V{c}}) \mid e_i = 0, \tilde{c}_i = 0) \notag \\
&= \sum_{e=e_\mathrm{min}}^{e_{\max}} \mathbb{P}(\mathcal{M}_d(\tilde{\V{c}}) \mid e_i=0, \tilde{c}_i = 0, w_\mathrm{H}({\V{e}}) = e) \notag \\
&\quad\times \mathbb{P}(w_\mathrm{H}({\V{e}}) = e \mid e_i = 0) \notag \\
&= \sum_{e=e_\mathrm{min}}^{e_{\max}} \sum_{q= q_\mathrm{min}^{(0)}}^{q_\mathrm{max}^{(0)}} \binom{d}{q} \binom{n-1-d}{e-q} p^e (1-p)^{n-1-e}\,,
\end{align}
\endgroup
where $q_\mathrm{min}^{(0)} \! = \! \max\ \!\{\left\lfloor d/2\right\rfloor \!+ \!1, \! \left\lceil (d\!+\!e\!-\!t)/2\rceil, \! e-(n-1-d)\right\}$ and $q_\mathrm{max}^{(0)}= \min\{d,e\}$, respectively.
Combining the above results, the analytical upper-bound of $F(p)$ is fully specified.

\begin{figure}[t]
    \centering
    \includegraphics[width=1.0\columnwidth]{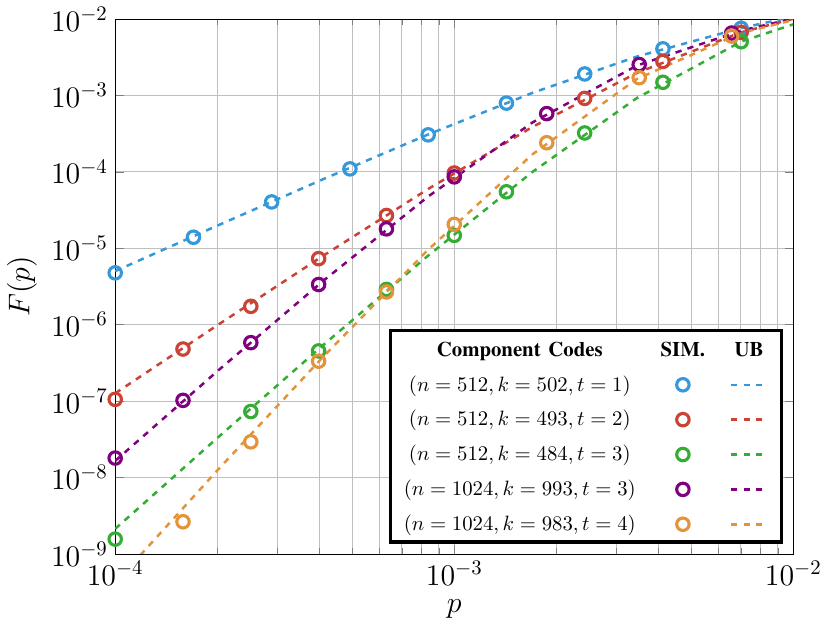}
    \caption{Analytical upper bound (UB) and Monte Carlo simulated points of the component-decoder update function $F(p)$ for the different eBCH component codes considered in the paper.}
    \label{fig:Analytical_transferFunction}
\end{figure}

To assess the tightness of the bound, Fig.~\ref{fig:Analytical_transferFunction} compares the analytical upper bounds with Monte Carlo simulated points of the update function $F(p)$ for all the \ac{eBCH} component codes considered in the \ac{GLDPC} codes design, reported in Table~\ref{tab:Design_Parameters}.
The results show that, in the region relevant to the waterfall analysis, the analytical curves (dashed lines) remain above the Monte Carlo results (circle markers) while being very close to them, confirming the tightness of the proposed upper bound.

The analytical recursion expressed in~\eqref{eq:pEvolution} exhibits a threshold-like behavior with respect to the initial channel crossover probability $p_0$: for sufficiently small $p_0$, the sequence ${p_\ell}$ converges to zero as $\ell \to \infty$, whereas for larger $p_0$ the recursion converges to a non-zero fixed point and a residual error probability remains.
This motivates the definition of the independence-based analytical threshold as
\begin{align}
p^{\star} = \text{sup} \{p_0>0, \text{s.t.} \lim_{\ell \to \infty} p_\ell = 0\}.
\end{align}
Thus, $p^{\star}$ denotes the largest initial channel error probability for which the independence-based recursion predicts convergence to zero error probability.
It is a threshold of the analytical surrogate and should not be interpreted as a rigorous decoding threshold of the finite-length deterministic fair-regular \ac{GLDPC}~code.

%%%%%%%%%%%%%%%%%%%%%%%%%%%%%%%%%%%%%%%%%%%%%%%%%%%%%%
%%%%%%%%%%%%%%%%%%%%%%%%%%%%%%%%%%%%%%%%%%%%%%%%%%%%%%

%%%%%%%%%%%%%%%%%%%%%%%%%%%%%%%%%%%%%%%%%%%%%%%%%%%%%%
%%%%%%%%%%%%%%%%%%%%%%%%%%%%%%%%%%%%%%%%%%%%%%%%%%%%%%
\section{Numerical Results}
\label{sec:NumericalResults}

\subsection{Simulation Setup}
\label{subsec:NumRes_SimSetup}
\begin{table}[t]
\caption{Designed GLDPC code constructions used in the simulations.}  
\label{tab:Design_Parameters}
\centering
\renewcommand{\arraystretch}{1.2}
\setlength{\tabcolsep}{2pt}
\begin{tabular}{lccc|c|ccc}
\toprule
\rowcolor[gray]{.95}
Parameter & D1 & D2 & D3 & D4 & D5 & D6 & D7 \\
\midrule
\multicolumn{8}{c}{\textit{Structure Parameters}} \\
\midrule
$N$ & \multicolumn{3}{c|}{$16384$} & $16384$ & \multicolumn{3}{c}{$32768$}\\

$M$ & \multicolumn{3}{c|}{$64$} & $96$ & \multicolumn{2}{c}{$96$} & $(64, 32)$\\

$n$ & \multicolumn{3}{c|}{$512$} & $512$ & \multicolumn{3}{c}{$1024$}  \\

$J$ & \multicolumn{3}{c|}{$2$} & $3$ & \multicolumn{3}{c}{$3$} \\

$V$ & \multicolumn{3}{c|}{$1024$} & $16384$ & \multicolumn{3}{c}{$32768$} \\

$v$ & \multicolumn{3}{c|}{$16$} & $1$ & \multicolumn{3}{c}{$1$} \\

$t$ & $1$ & $2$ & $3$ & $2$ & $3$ & $4$ & $(3,4)$ \\
\midrule
\multicolumn{8}{c}{\textit{Nominal Design Values}} \\
\midrule
$(N-K)_\mathrm{n}$ & 640 & 1216 & 1792 & 1824 & 2976 & 3936 & 3296\\
$R_\mathrm{n}$ & $0.961$ & $0.926$ & $0.891$ & $0.889$ & $0.909$ & $0.880$ & $0.899$ \\
$\mathrm{OH}_\mathrm{n}~[\%]$ & $4.07$ & $8.02$ & $12.28$ & $12.53$ & $9.99$ & $13.65$ & $11.18$ \\
\midrule
\multicolumn{8}{c}{\textit{Actual Code Values}} \\
\midrule
$N-K$ & $630$ & $1197$ & $1764$ & $1786$ & $2914$ & $3854$ & $3234$ \\
$R$ & $0.962$ & $0.927$ & $0.892$ & $0.891$ & $0.911$ & $0.882$ & $0.901$ \\
$\mathrm{OH}~[\%]$ & $4.00$ & $7.88$ & $12.07$ & $12.23$ & $9.76$ & $13.33$ & $10.95$ \\
\bottomrule
\end{tabular}
\end{table}
Table~\ref{tab:Design_Parameters} summarizes the high-rate \ac{GLDPC} code configurations considered in the numerical results.
Each design is characterized by its \ac{VN}-\ac{CN} connectivity, component-code classes, and target block length.
The structured connectivity is determined by the fair-regular construction introduced in Section~\ref{subsec:FairRegular}, whereas the component codes specify the local constraints at the \acp{CN}.
All considered designs employ \ac{eBCH} component codes, whose length $n$ and error-correction capability $t$ are reported in the table.
For these codes, the minimum distance is $d_{\min}=2t+2$, resulting in $d_{\min}\in\{4,6,8,10\}$ for $t\in\{1,2,3,4\}$, respectively.
For designs with multiple component-code classes, the corresponding \ac{eBCH} parameters are reported separately.
For each design, the table reports the main code parameters: \ac{GLDPC} codeword length $N$, number of \acp{CN} $M$, \ac{VN} degree $J$, number of \acp{SVN} $V$, and \ac{SVN} size $v$.
In addition, both nominal and actual coding metrics are provided.
The nominal redundancy $(N-K)_\mathrm{n}$, nominal code rate $R_{\mathrm n}$, and nominal overhead $\mathrm{OH}_{\mathrm n}$ are obtained from the design equations presented in Section~\ref{sec:Design}, whereas the actual redundancy $N-K$, code rate $R$, and overhead $\mathrm{OH}$ are evaluated after graph construction, accounting for possible rank deficiencies or structural dependencies in the resulting parity-check matrix.

Designs D1-D3 adopt a \ac{VN} degree $J=2$ and a block length $N=16384$.
They are primarily used to empirically assess both the independence-based waterfall prediction and the finite-length error floor analysis by comparison with Monte Carlo simulation results.
Moreover, these designs allow to isolate the impact of the component code error-correction capability $t \in \{1,2,3\}$ under fixed graph connectivity.
Design D4 increases the \ac{VN} degree to $J=3$ while maintaining the same block length, thereby enabling a direct comparison between degree-$2$ and degree-$3$ graph structures.
Designs D5-D7 consider the larger block length $N=32768$ and are used to investigate the benefits of longer codewords together with stronger or heterogeneous component code classes.
Unless otherwise stated, decoding is performed via conventional \ac{iBDD} with a maximum number of \ac{HD} iterations equal to $\ell_{\max}=5$.
All Monte Carlo simulations are carried out over the \ac{bi-AWGN} channel with \ac{BPSK} modulation.
The received soft values are first hard-quantized and then processed by the \ac{iBDD} decoder, which only exploits binary decisions.
This setup reflects practical \ac{HD} optical receivers, while still allowing performance to be reported in the form of \ac{BER} versus~$E_\mathrm{b}/N_0$.

\subsection{Performance Evaluation}
\label{subsec:NumRes_PerformanceEval}

\subsubsection{Validation of analytical tools}
\begin{figure}[t]
    \centering
    \includegraphics[width=1.0\columnwidth]{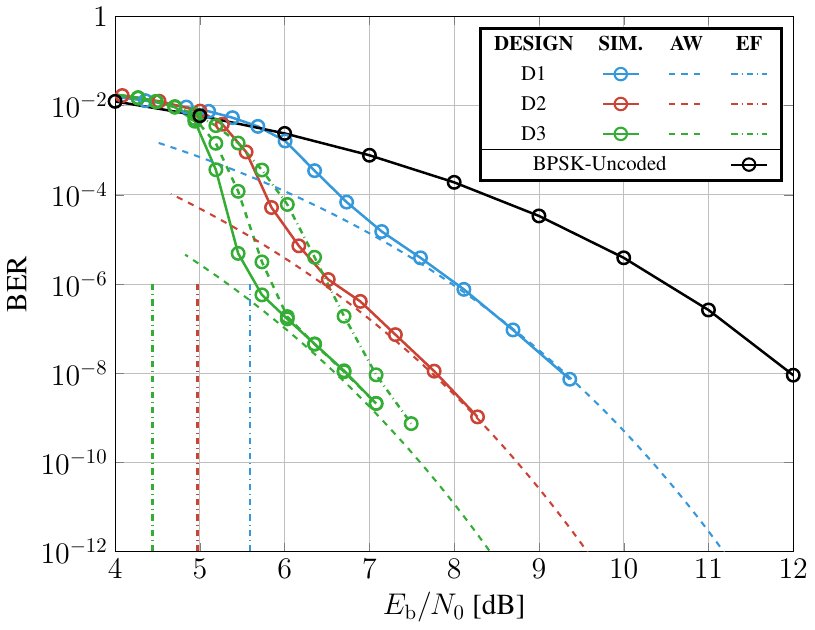}
    \caption{BER performance of fair-regular \ac{GLDPC} codes D1-D3 with identical graph parameters $(N,M,n,J,V,v)=(16384,64,512,2,1024,16)$ and different \ac{eBCH} component code correction capabilities $t$.
    Simulated \ac{BER} curves are compared with uncoded \ac{BPSK}, analytical waterfall (AW) surrogate predictions, and finite-length error floor (EF) estimates.
    The considered setup satisfies $v>\max_j(t_j)$, which leads to a pronounced error floor.
    Additional curves for D3 with $\ell_{\max}=1$ (green circle dash-dot) and $\ell_{\max}=2$ (green circle dashed) iterations illustrate the convergence behavior of \ac{iBDD}.}
    \label{fig:Plot_SimVsLB_vGreaterThant_Vart}
\end{figure}
\begin{figure*}[t]
    \centering
    \includegraphics[width=1.0\textwidth]{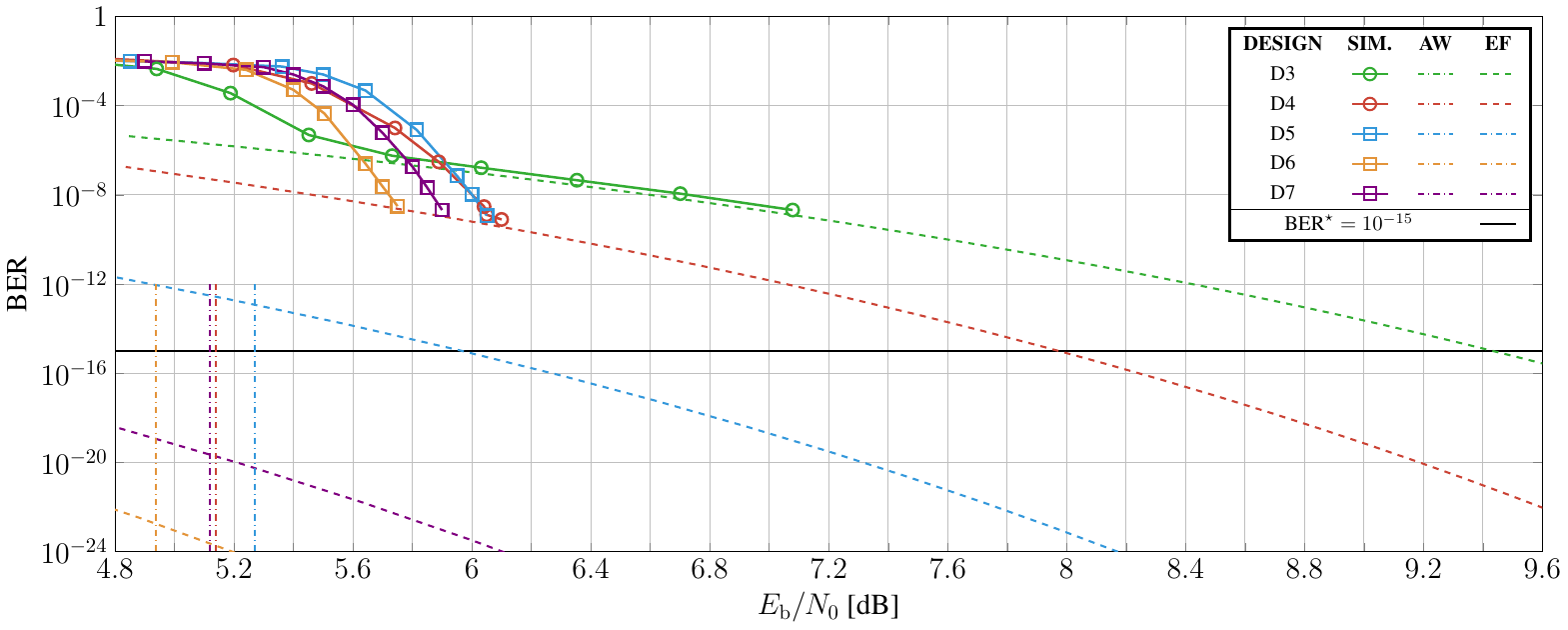}
    \caption{BER performance of the considered fair-regular \ac{GLDPC} code designs for $J=2$ and $J=3$. Monte Carlo simulation results are compared with the finite-length error floor (EF) predictions and the analytical waterfall (AW) surrogate predictions.
    Dashed lines denote the error floor estimates derived in Section~\ref{sec:ErrorFloorAnalysis} under miscorrection-free \ac{iBDD}, dash-dotted lines denote the waterfall predictions obtained from the independence-based analytical recursion in Section~\ref{sec:AsymptoticWaterfallAnalysis}, whereas markers indicate simulated results. 
    The horizontal line marks the target \ac{BER} of $10^{-15}$ adopted for \ac{NCG} evaluation.}
    \label{fig:Simulation_EF_J2vsJ3}
\end{figure*}
Fig.~\ref{fig:Plot_SimVsLB_vGreaterThant_Vart} compares the first three code designs in Table~\ref{tab:Design_Parameters} (D1-D3) to validate the analytical tools developed in Section~\ref{sec:ErrorFloorAnalysis} and Section~\ref{sec:AsymptoticWaterfallAnalysis}, and to isolate the impact of the adopted component codes.
In this experiment, all \ac{GLDPC} codes share the same structural parameters, namely block length $N=16384$, number of \acp{CN} $M=64$, component code length $n=512$, and \ac{VN} degree $J=2$. Hence, the only varying parameter is the error-correction capability of the \ac{eBCH} component code.

As expected, increasing $t$ enhances the local decoding capability of each \ac{CN} and yields a clear performance improvement in both the waterfall and the error floor regions.
In the waterfall region, stronger component codes shift the operating point toward lower $E_\mathrm{b}/N_0$, consistently with the performance trends predicted by the independence-based analysis of Section~\ref{sec:AsymptoticWaterfallAnalysis}.
The analytical thresholds should be interpreted as predictions of the surrogate recursion, and not as rigorous asymptotic thresholds for the deterministic fair-regular \ac{GLDPC} code family.
The comparison with the Monte Carlo results shows that the surrogate correctly captures the waterfall trends and the relative performance ordering, even for the relatively small lifting factor $Q=32$ considered here. 
In the low-\ac{BER} region, larger values of $t$ reduce the contribution of harmful stalling sets, resulting in a lower error floor.
We remark that the observed behavior closely matches the analytical bound derived in Section~\ref{sec:ErrorFloorAnalysis}.
This confirms that, although the analytical bound is derived under the miscorrection-free \ac{iBDD} assumption, its close agreement with the simulated performance under conventional \ac{iBDD} indicates that problematic miscorrections remain negligible in this regime.
This gain, however, is obtained at the expense of spectral efficiency.
Indeed, moving from D1 to D3 requires progressively more redundancy, which increases the overhead from $\text{OH}=4.07\%$ (D1) to $\text{OH}=12.28\%$ (D3).

Fig.~\ref{fig:Plot_SimVsLB_vGreaterThant_Vart} also reports the performance of code D3 when the decoder is limited to $\ell_{\max}=1$ and $\ell_{\max}=2$ \ac{iBDD} iterations.
Most of the achievable gain is already obtained within the first few iterations, with convergence reached for $\ell_{\max}=5$, which is the default value adopted throughout the paper.
This is particularly relevant for high-speed optical communication systems, where decoding latency is a crucial constraint.

\subsubsection{Performance of fair-regular GLDPC designs}

Fig.~\ref{fig:Simulation_EF_J2vsJ3} reports the main performance results of the proposed fair-regular \ac{GLDPC} codes and highlights the joint impact of component code selection and block length on both the waterfall and error floor regions.
As previously observed, increasing the component code error correction capability $t$ improves performance of the degree-$2$ designs D1-D3. However, even the strongest such configuration still exhibits an error floor that is far too high for applications targeting \ac{BER} levels around $10^{-15}$.
The fundamental reason is that these designs operate in the regime $v>\max_j(t_j)$, where low-weight stalling sets remain comparatively likely and therefore dominate the low-\ac{BER} region.
As a consequence, although D3 shows a clear waterfall improvement, its achievable \ac{NCG} at $\mathrm{BER}^{\star}=10^{-15}$ is limited to approximately $5.5$~dB, corresponding to $(E_\mathrm{b}/N_0)^{\star}\approx 9.5$~dB due to the onset of the error floor.

To overcome this limitation, the graph connectivity must be redesigned so as to move toward the more favorable regime $v \le \max_j(t_j)$.
This motivates the transition from $J=2$ to $J=3$, which increases the Tanner-graph connectivity and modifies the structure of the dominant stalling sets.
In particular, minimum-weight unsolvable error patterns require a larger number of involved \acp{VN} and more simultaneously failing component constraints, thereby significantly reducing their occurrence probability.
The first design following this principle is D4.
In order to maintain an overall code rate close to that of D3, D4 adopts degree $J=3$ while reducing the component code error correction capability from $t=3$ to $t=2$.
The block length remains unchanged at $N=16384$, while the number of \acp{CN} slightly increases.
The resulting performance confirms the expected tradeoff: compared with D3, D4 achieves a substantial reduction of the error floor thanks to the improved graph structure, while its waterfall region is moderately degraded due to the weaker component code.
Nevertheless, despite the clear improvement, D4 still remains above the target \ac{BER} for practically attractive operating points, yielding an \ac{NCG} of about $7.0$~dB at $\mathrm{BER}^{\star}=10^{-15}$, corresponding to $(E_\mathrm{b}/N_0)^{\star}\approx 8$~dB.
We remark that, for D4, the analytical error floor estimation should be interpreted as an approximation, since not all possible combinations of one \ac{CN} per type correspond to a non-empty \ac{VN} set.
Consequently, the analytical estimate may lie below the simulated error floor.
Nevertheless, as shown by the numerical results, the analytical estimate provides a good approximation of the observed error floor over the considered $E_\mathrm{b}/N_0$ range.

A second design dimension is then explored, namely the codeword length.
Increasing the block size from $N=16384$ to $N=32768$ naturally increases transmission latency, since larger blocks must be encoded and decoded, but it also enables stronger global coding performance.
This motivates the analysis of designs D5-D7, all based on $J=3$ and $n=1024$.
Design D5 employs a single component code class with correction capability $t=3$.
It already provides a markedly lower error floor, close to the target region, while preserving a very small overhead of only $9.76\%$ ($R=0.911$).
At $\mathrm{BER}^{\star}=10^{-15}$, D5 achieves an estimated \ac{NCG} of approximately $8.8$~dB.
The \ac{NCG} is estimated by averaging the last simulated point and extrapolating the resulting waterfall-region trend to $\mathrm{BER}^{\star}$, since direct Monte Carlo estimation at such low BER values would require prohibitively long simulation times.
In practice, hardware-accelerated simulations are typically required for performance evaluation in this regime.
Design D6 further strengthens the component codes to $t=4$.
This leads to an even lower error floor and an improved waterfall region, resulting in the best overall gain among the considered schemes, with an \ac{NCG} of approximately $9.08$~dB at the target \ac{BER}.
The price to pay is a larger redundancy, corresponding to an overhead of $13.33\%$.

Finally, design D7 provides an effective compromise through the use of multiple component code classes.
Specifically, the type-$1$ \acp{CN}, i.e., the first component decoders processed within each \ac{iBDD} iteration, employ the stronger $t=4$ component codes, whereas the remaining type-$2$ and type-$3$ \acp{CN} employ $t=3$ component codes.
This asymmetric assignment can be particularly beneficial because the decoding process is sequential: early successful corrections performed by the strongest component decoders can provide cleaner inputs to the subsequent \acp{CN} processed within the same iteration.
As a result, the stronger type-$1$ constraints may promote the propagation of error corrections through the decoding sequence, facilitating the convergence of the remaining weaker component decoders.
The resulting architecture achieves an error floor comfortably below the target while preserving a waterfall performance close to D6, despite the lower average redundancy.
Overall, D7 attains an \ac{NCG} of approximately $9.01$~dB with an overhead of only $10.95\%$, representing an attractive tradeoff between coding gain and redundancy.

\subsubsection{NCG comparison with optical FEC codes}
\begin{table}[t]
\caption{\ac{NCG} comparison at $\mathrm{BER}^{\star}=10^{-15}$ between the proposed fair-regular \ac{GLDPC} designs and representative reference optical \ac{FEC} schemes. The last column reports the gap with respect to the theoretical HD and SD decoding limits.}
\label{tab:NCG_OH_CodeDesign}
\centering
\renewcommand{\arraystretch}{1.15}
\setlength{\tabcolsep}{1pt}
\begin{tabular}{lccccc}
\toprule
\rowcolor[gray]{.95}
Code & Type & OH [\%] & NCG [dB] & Limit [dB] & Gap [dB] \\
\midrule
\multicolumn{6}{c}{\textit{Reference Optical FEC Schemes}}\\
\midrule
KR4     & HD & 2.72  & 5.59  & 8.99~(HD)  & 3.40\\%~\cite{IEEE8023bj}
HOSC    & HD & 4.17  & 8.69  & 9.44~(HD)  & 0.75\\%~\cite{Shehadeh2025:HOSC}
KP4     & HD & 5.83  & 6.63  & 9.82~(HD)  & 3.19\\%~\cite{IEEE8023bj}
SC      & HD & 6.69  & 9.41  & 9.97~(HD)  & 0.56\\%~\cite{Smith2012:StaircaseCodes}
KP4+SPC & HD/SD & 11.13 & 7.96  & 11.81~(SD) & 3.85\\%~\cite{Lentner2023:ConcatenatedKp4}
KP4+eH  & HD/SD & 12.89 & 9.21  & 12.01~(SD) & 2.80\\
CFEC    & HD/SD & 14.76 & 10.34 & 12.20~(SD) & 1.86\\%~\cite{Wang2023:CFEC_OFEC}
OFEC    & HD/SD & 15.34 & 11.21 & 12.26~(SD) & 1.05\\%~\cite{Wang2023:CFEC_OFEC}
\midrule
\multicolumn{6}{c}{\textit{Proposed Fair-Regular GLDPC Designs}}\\
\midrule
D5   & HD    & 9.76  & 8.80 & 10.47~(HD) & 1.67 \\
D7   & HD    & 10.95 & 9.01 & 10.62~(HD) & 1.61 \\
D6   & HD    & 13.33 & 9.08 & 10.86~(HD) & 1.78 \\
\bottomrule
\end{tabular}
\end{table}
After quantifying the \ac{NCG} achieved by the proposed fair-regular \ac{GLDPC} designs, we now position them with respect to representative \ac{FEC} solutions adopted in high-speed optical communication systems.
Table~\ref{tab:NCG_OH_CodeDesign} reports a comparison at the target \ac{BER} of $10^{-15}$, including both practical schemes from the literature and the theoretical decoding limits at the same overhead.
Since the considered \ac{FEC} schemes operate at different coding rates, their absolute \ac{NCG} values should be interpreted together with their coding overhead.
To this end, Table~\ref{tab:NCG_OH_CodeDesign} also reports the corresponding theoretical \ac{HD} or \ac{SD} \ac{NCG} limit and the gap between the achieved \ac{NCG} and this limit.
The theoretical limits are obtained from the capacity of the binary-input AWGN channel under either hard- or soft-decision reception, respectively, and therefore represent the fundamental performance bounds associated with the corresponding decoding paradigm.
The gap to the corresponding limit therefore provides a rate-aware measure of the efficiency of each scheme.
We first consider conventional end-to-end \ac{HD} \ac{FEC} schemes, such as KR4 and KP4.
These solutions are highly attractive from an implementation viewpoint, since they rely on low-complexity \ac{HD} decoding and require very limited redundancy, with overheads of only $2.72\%$ and $5.83\%$, respectively.
However, their achievable \ac{NCG} remains moderate, equal to $5.59$~dB for KR4 and $6.63$~dB for KP4, and they still operate several decibels away from the corresponding \ac{HD} theoretical limits.

A substantial performance improvement can be obtained through staircase-based \ac{HD} coding schemes, such as \acp{SC} and \acp{HOSC}.
These architectures still maintain relatively small overheads ($6.69\%$ and $4.17\%$, respectively), while significantly increasing the \ac{NCG} to $9.41$~dB and $8.69$~dB.
Consequently, their gap to the \ac{HD} limit becomes rather small.
Nevertheless, such staircase-based codes typically rely on sliding-window decoding, which may increase decoding latency and architectural complexity compared with block-based solutions.
We then consider hybrid and concatenated schemes combining \ac{HD} and \ac{SD} processing.
Examples include CFEC and OFEC, which achieve strong gains of $10.34$~dB and $11.21$~dB, respectively, at overheads around $15\%$.
Similarly, concatenated solutions such as KP4+SPC and KP4+eH operate in an overhead region closer to the one targeted in this work (roughly $11\%$--$13\%$), with \ac{NCG} values of $7.96$~dB and $9.21$~dB.
These schemes benefit from reliability information and therefore approach the \ac{SD} limits more closely, at the price of increased decoder complexity, power consumption, and implementation cost.

We now turn to the proposed fair-regular \ac{GLDPC} codes, all decoded through low-complexity \ac{iBDD} with purely \ac{HD} messages.
Among the considered designs, the most attractive operating points are obtained by the longer-block ones.
Design D5 and D6 reach a gap of only $1.67$~dB and $1.78$~dB to the \ac{HD} limit, respectively.
Finally, D7 achieves the smallest \ac{NCG} gap to the \ac{HD} limit among the proposed solutions, providing a more favorable tradeoff between coding overhead and \ac{NCG}.
Thanks to its multi-class component codes allocation, it achieves the \ac{NCG} gap of $1.61$~dB.

\subsubsection{Discussion on latency with HD optical FEC codes}
\begin{table}[t]
\centering
\caption{Transmission buffering latency comparison for representative \ac{HD} optical \ac{FEC} schemes at an aggregate coded bit rate $B_r=1.6$~\textnormal{Tb/s}.
For staircase-like schemes, the effective buffered length includes the decoding window $W$.}
\label{tab:Latency}
\renewcommand{\arraystretch}{1.18}
\setlength{\tabcolsep}{5pt}
\begin{tabular}{lcc}
\toprule
\rowcolor[gray]{.95}
Code Scheme & Buffered Bits, $N$ & $\Omega_{\mathrm{tx}}$ [ns] \\
\midrule

\multicolumn{3}{c}{\textit{Short Block HD-FEC}}\\
\midrule
KR4 & $5280$ & $3.30$ \\
KP4 & $5440$ & $3.40$ \\

\midrule
\multicolumn{3}{c}{\textit{Proposed Fair-Regular GLDPC}}\\
\midrule
% D3--D4 ($N=16384$) & $16384$ & $10.24$ \\
D5-D7 ($N=32768$) & $32768$ & $20.48$ \\

\midrule
\multicolumn{3}{c}{\textit{Window-Based HD-FEC}}\\
\midrule
SC~\cite{Smith2012:StaircaseCodes} & $\sim 1.83\times10^6$ & $\sim 1142.4$ \\
SC ($N_\mathrm{B} = 16384,\,W=5$) & $\sim 8.19\times10^4$ & $\sim51.2$ \\
SC ($N_\mathrm{B} = 16384,\,W=7$) & $\sim 1.14\times10^5$ & $\sim71.7$ \\

\bottomrule
\end{tabular}
\end{table}

To complete the comparison among the considered coding schemes, we now discuss latency considerations, which are of primary importance in high-speed optical communication systems.
Since the proposed fair-regular \ac{GLDPC} codes employ purely \ac{HD} iterative decoding, we focus on representative \ac{HD}-oriented optical \ac{FEC} schemes.
Schemes involving \ac{SD} processing are not directly comparable from a latency and implementation viewpoint, as they require reliability quantization, additional memory, and significantly more complex decoding operations.
In practical systems, a first-order contribution to latency is the time required to receive the coded bits before decoding can start.
For a block code of length $N$, the transmission latency can be approximated as
\begin{align}
\Omega_{\mathrm{tx}} = \frac{N}{B_r},
\end{align}
where $B_r$ denotes the received coded bit rate.
Assuming a representative aggregate throughput of $B_r = 1.6$~Tb/s, corresponding, for instance, to four parallel $400$~Gb/s lanes, the resulting transmission delays are reported in Table~\ref{tab:Latency}.
Conventional short-block \ac{HD}-\ac{FEC} schemes such as KR4 and KP4 operate on codewords of approximately $5.3$-$5.4 \times 10^3$ bits, resulting in transmission latencies around $3.3$-$3.4$ ns.
The proposed fair-regular \ac{GLDPC} constructions operate with moderate block lengths.
In particular, designs D5-D7 employ $N=32768$, corresponding to $\Omega_{\mathrm{tx}} \approx 20.48$ ns.
Although larger than KR4/KP4, these values remain small in absolute terms and are fully compatible with low latency optical links.

A relevant comparison is with staircase-based \ac{HD}-\ac{FEC} schemes.
Unlike block codes, staircase decoding relies on a sliding-window architecture spanning multiple consecutive blocks.
The original G.709-compatible \ac{SC} proposed in~\cite{Smith2012:StaircaseCodes} employs BCH$(1023,993)$ component codes, resulting in staircase blocks of approximately $N_\mathrm{B}=261120$ coded bits.
We also consider a second example based on BCH$(255,239)$ component codes, for which a single staircase block contains approximately $N_\mathrm{B}=128 \times 128 = 16384$ coded bits.
Although this configuration may exhibit some performance degradation compared with the original design, it enables a fair comparison in terms of block size, since it is comparable to one full proposed \ac{GLDPC} codeword with $N=16384$.

An important architectural difference is that the proposed fair-regular \ac{GLDPC} codes can start decoding immediately after the reception of a single codeword, whereas staircase decoding requires buffering an entire decoding window before iterative processing can begin.
In practical staircase decoders, window sizes typically range between $W=5$ and $W=10$.
As regards the original \ac{SC} implementation with decoding window $W=7$, the total buffered size is approximately $1.83\times10^6$ bits, corresponding to a transmission buffering latency of about $1142.2$~ns at $1.6$~Tb/s.
For the block-compatible \ac{SC} example, even for moderate window sizes such as $W=5$ and $W=7$, the buffered sizes increase to approximately $8.19\times10^4$ and $1.14\times10^5$ bits, corresponding to latencies of about $51.2$ ns and $71.7$ ns, respectively.

The actual decoding latency is not reported as a fixed numerical quantity in Table~\ref{tab:Latency}, since it is strongly implementation-dependent and varies with the decoder architecture, degree of parallelism, pipelining, memory organization, and clock frequency.
In particular, the latency $\Omega_\mathrm{dec}^{(c)}$ of a component-code \ac{BDD} operation depends on the specific decoder implementation, while the overall decoding latency further depends on the decoding architecture and the number of iterations.
For the proposed fair-regular \ac{GLDPC} codes, for example, under the \ac{iBDD} mechanism considered in this work, \acp{CN} of the same type can be decoded in parallel, whereas the $J$ \ac{CN} types are processed sequentially.
Thus, assuming a fixed number $\ell_\mathrm{
max}$ of iterations, the decoding latency can be expressed as $\Omega_\mathrm{dec} = \ell_{\max} \sum_{j=1}^J \Omega_\mathrm{dec}^{(c,j)}$, where $\Omega_\mathrm{dec}^{(c,j)}$ denotes the latency of the component-code decoder associated with CN type $j$.
Similarly, for staircase codes, $\Omega_\mathrm{dec}$ depends on the adopted window architecture and iterative decoding schedule.
In general, practical high-throughput optical \ac{FEC} decoders employ substantial parallelization and pipelining to meet stringent throughput and latency requirements~\cite{Fougstedt2019:VLSI}. 
Moreover, efficient table-lookup implementations have been proposed for low-complexity component-code decoding in staircase codes~\cite{Smith2012:StaircaseCodes}, enabling very short component-decoding latency.
We therefore retain $\Omega_\mathrm{tx}$ as the quantitative, hardware-independent latency metric.
These results highlight a key advantage of the proposed fair-regular \ac{GLDPC} constructions: even the longest considered designs ($N=32768$) require significantly smaller buffering than practical staircase decoding windows, while maintaining a purely \ac{HD} low-complexity decoding architecture.

%%%%%%%%%%%%%%%%%%%%%%%%%%%%%%%%%%%%%%%%%%%%%%%%%%%%%%
%%%%%%%%%%%%%%%%%%%%%%%%%%%%%%%%%%%%%%%%%%%%%%%%%%%%%%

%%%%%%%%%%%%%%%%%%%%%%%%%%%%%%%%%%%%%%%%%%%%%%%%%%%%%%
%%%%%%%%%%%%%%%%%%%%%%%%%%%%%%%%%%%%%%%%%%%%%%%%%%%%%%
\section{Conclusions}
\label{sec:Conclusions}
Overall, the proposed fair-regular \ac{GLDPC} constructions exhibit a highly favorable tradeoff among \ac{NCG}, very low \ac{BER} performance, and transmission latency, making them strong candidates for next-generation high-speed optical interconnects and data-center links.
In particular, the results show that carefully designed high-rate \ac{GLDPC} codes can approach the performance of reference optical \ac{FEC} schemes while preserving the architectural simplicity and efficiency of \ac{iBDD}-based decoding.
Several research directions naturally emerge from this work.
A first important topic concerns the development of low-complexity encoding architectures for high-rate \ac{GLDPC} codes, possibly leveraging the structure of the component encoders, similarly to efficient implementations used for product codes and \acp{SC}.
A second direction is the investigation of enhanced decoding algorithms capable of improving the waterfall region while maintaining limited complexity and latency.
Promising approaches include reliability-aided \ac{HD} decoding, soft-assisted iterative decoding, or lightweight Chase-type soft-input hard-output strategies, where a small amount of soft information is exploited without incurring the full cost of conventional \ac{SD} decoding.

%%%%%%%%%%%%%%%%%%%%%%%%%%%%%%%%%%%%%%%%%%%%%%%%%%%%%%
%%%%%%%%%%%%%%%%%%%%%%%%%%%%%%%%%%%%%%%%%%%%%%%%%%%%%%

\appendices

\section{PROOF OF LEMMA 2}

Let
\begin{align}
P=\prod_{j=1}^{J} |\mathcal{S}_j|.
\end{align}
Since every $J$-tuple of \acp{CN} is connected to exactly $v$ \acp{VN}, label those $v$ \acp{VN} by $1,\ldots,v$. 
This partitions the \ac{VN} set into $v$ layers, each containing exactly one \ac{VN} for every possible $J$-tuple of selected \acp{CN}.
We first consider a single layer. 
Index the \acp{CN} of type $j$ by $0,1,\ldots,|\mathcal{S}_j|-1$.
Each \ac{VN} in the layer can then be represented uniquely by a $J$-tuple
\begin{align}
(x_1,\ldots,x_J),
\qquad
x_j\in\{0,\ldots,|\mathcal{S}_j|-1\}.
\end{align}
Define $L=\operatorname{lcm} \left( |\mathcal{S}_1|,\ldots,|\mathcal{S}_J| \right)$.
Starting from any \ac{VN} $(x_1,\ldots,x_J)$, construct the sequence $(x_1+k,\ldots,x_J+k)$, with $k=0,\ldots,L-1$, where the $j$-th coordinate is taken modulo $|\mathcal{S}_j|$.
The $L$ \acp{VN} in this sequence are all distinct. 
Moreover, for each type $j$, the $j$-th coordinate cycles through all \acp{CN} in $\mathcal{S}_j$, so every \ac{CN} of type $j$ appears exactly $L/|\mathcal{S}_j|$ times.
Choosing any \ac{VN} not yet included and repeating the same construction partitions the whole layer into disjoint sequences of length $L$.
Hence, for any integer $0\le r<P$, one can select $r$ \acp{VN} from a single layer such that, for every type $j$, each \ac{CN} in $\mathcal{S}_j$ is connected to either
\begin{align}
\left\lfloor \frac{r}{|\mathcal{S}_j|}\right\rfloor
\qquad \text{or} \qquad
\left\lceil \frac{r}{|\mathcal{S}_j|}\right\rceil
\end{align}
selected \acp{VN}.
Now write
\begin{align}
e=qP+r, \qquad 0\le r<P,
\end{align}
with $0\le q\le v$.
Select all \acp{VN} from $q$ complete layers, and then select $r$ \acp{VN} from one additional layer according to the construction above.
Each complete layer contributes exactly $P/|\mathcal{S}_j|$ selected \acp{VN} to every \ac{CN} of type $j$. 
Therefore, the $q$ complete layers contribute $q P/|\mathcal{S}_j|$ selected \acp{VN} uniformly to every \ac{CN} of that type.
The remaining $r$ selected \acp{VN} contribute either $\left\lfloor r / |\mathcal{S}_j|\right\rfloor$ or $\left\lceil r / |\mathcal{S}_j| \right\rceil$.
Since $P$ is divisible by $|\mathcal{S}_j|$, the total number of selected \acp{VN} connected to each \ac{CN} of type $j$ is therefore either
\begin{align}
\left\lfloor \frac{e}{|\mathcal{S}_j|}\right\rfloor \quad \text{or} \quad \left\lceil \frac{e}{|\mathcal{S}_j|}\right\rceil.
\end{align}
Thus, for every type $j$, the selected \acp{VN} can be distributed among the \acp{CN} in $\mathcal{S}_j$ as uniformly as possible.

%%%%%%%%%%%%%%%%%%%%%%%%%%%%%%%%%%%%%%%%%%%%%%%%%%%%%%
%%%%%%%%%%%%%%%%%%%%%%%%%%%%%%%%%%%%%%%%%%%%%%%%%%%%%%
%%%%%%%%%%%%%%%%%%%%%%%%%%%%%%%%%%%%%%%%%%%%%%%%%%%%%%
%%%%%%%%%%%%%%%%%%%%%%%%%%%%%%%%%%%%%%%%%%%%%%%%%%%%%%

%%%%%%%%%%%%%%%%%%%%%%%%%%%%%%%%%%%%%%%%%%%%%%%%%%%%%%
%%%%%%%%%%%%%%%%%%%%%%%%%%%%%%%%%%%%%%%%%%%%%%%%%%%%%%
\balance
\bibliographystyle{IEEEtran}
\bibliography{Files/IEEEabrv,Files/StringDefinitions,Files/StringDefinitions2,Files/refs}
%%%%%%%%%%%%%%%%%%%%%%%%%%%%%%%%%%%%%%%%%%%%%%%%%%%%%%
%%%%%%%%%%%%%%%%%%%%%%%%%%%%%%%%%%%%%%%%%%%%%%%%%%%%%%

\end{document}